\documentstyle[12pt]{article}

\def\ba{\begin{array}}
\def\ea{\end{array}}
\def\be{\begin{equation}\begin{array}{l}}
\def\ee{\end{array}\end{equation}}
\def\bea{\begin{equation}\begin{array}{l}}
\def\eea{\end{array}\end{equation}}
\def\f{\frac}
\def\om{\omega}
\def\Om{\Omega}
\def\we{\wedge}
\def\d{\delta}

\def\ep{\epsilon}

\def\t{\tau}

\def\c{\cite}
\def\D{\Delta}
\def\a{\alpha}
\def\b{\beta}
\def\n{\nabla}
\def\pa{\partial}
\begin{document}
\begin{flushright}
 May 30, 2001
\end{flushright}

\begin{center}
{\large \bf
DIFFERENCE DISCRETE VARIATIONAL PRINCIPLE,\\
 EULER-LAGRANGE COHOMOLOGY AND\\

\vskip 1.5mm
SYMPLECTIC, MULTISYMPLECTIC STRUCTURES}

\vspace{4mm}

H.Y. Guo,  
\quad Y.Q. Li,
\quad
  K. Wu {\footnote{Email: hyguo@itp.ac.cn; qylee@itp.ac.cn; wuke@itp.ac.cn.}}\\
Institute of Theoretical    Physics,    Academia   Sinica\\
P.O. Box 2735, Beijing 100080, China. 
\vskip 1.0mm
and\\
\vskip 1.0mm
S.K. Wang {\footnote{Email: wsk@amath5.ami.ac.cn.}}\\
Institute of Applied Mathematics\\
Academy of Mathematics and Systems Science, Academia Sinica\\
P.O. Pox 2734, Beijing 100080, China.

\vskip 2mm 
\end{center}
 \vskip 2mm \noindent
{\centerline {\sc Abstract}}
{\small 
We study the difference discrete variational principle  in the
framework of multi-parameter differential approach by regarding
the forward difference as an entire geometric object in view of
noncomutative differential geometry. By virtue of this variational
principle, we get the difference discrete Euler-Lagrange equations
and canonical ones for the difference discrete versions of the
classical mechanics and classical field theory. We also explore
the difference discrete versions for the Euler-Lagrange cohomology
and apply them to the symplectic or multisymplectic geometry
and their preserving properties in both 
Lagrangian and Hamiltonian formalism. 
In terms of the difference discrete Euler-Lagrange cohomological
concepts, we show that the symplectic or multisymplectic geometry
and their difference discrete structure preserving properties can
always be established not only in the solution spaces of the
discrete Euler-Lagrange/canonical equations derived by the
difference discrete variational principle but also in the function
space in each case if and only if the relevant closed
Euler-Lagrange cohomological conditions are satisfied. We also
apply the difference discrete variational principle and
cohomological approach directly to
the symplectic and multisymplectic algorithms.}

\vskip 5mm
{\parindent=0mm {\large \sc contents}\it

1. Introduction

2.  Variational  and difference discrete variational principle in
multi-parameter differential approach

2.1. Variational and  difference discrete variational principle in
 continuous and difference discrete classical 
 mechanics

2.2. Variational  and difference discrete variational principle in
  continuous and discrete classical 
  field theory

3. Euler-Lagrange cohomology, symplectic structure preserving
property in continuous and difference discrete classical mechanics

3.1. Euler-Lagrange cohomology, symplectic structure preserving
property in continuous classical mechanics

3.2. Discrete Euler-Lagrange cohomology and symplectic structure
preserving property in difference discrete classical mechanics


4. Euler-Lagrange cohomology, multisymplectic structure preserving
property in continuous and difference discrete classical field
theory

4.1. Euler-Lagrange cohomology, multisymplectic structure
preserving property in continuous  classical field theory

4.2. Discrete Euler-Lagrange cohomology and multisymplectic
structure preserving property in difference discrete classical
field theory


5. Difference discrete variational principle and discrete
Euler-Lagrange cohomological approach to symplectic and
multisynplectic Algorithms

5.1. Difference discrete variational principle and discrete
cohomological approach to symplectic algorithm

5.2. Difference discrete variational principle and discrete
cohomological approach to multisymplectic algorithm

6. Concluding remarks

Appendix: An noncommutative differential calculus on regular
lattice.

Acknowledgements

 References}
\section{Introduction}

$\quad$ It is well known that the symplectic 
structure plays  crucially important role in the both Lagrangian
and Hamiltonian formalism for classical mechanics
\c{ALD78}\c{AM78}. On the other hand, the multisymplectic
structure plays also very important role in the Lagrangian and
Hamiltonian formalism for classical field theories
\c{BSF88}\c{TB97}\c{GIM97}\c{MS98}\c{MPS98}. Specially, in the
computational science, they are extremely important in the
symplectic and multisymplectic algorithms for the finite
dimensional Hamiltonian
systems 
\c{KF84}\c{SC94} and infinite-dimensional 
\c{TB97}\c{MPS98} respectively. These algorithms are  quite
powerful  and successful  in  numerical calculations of  the
relevant  systems in comparison with  other various
non-symplectic/multisymplectic numerical  schemes since the
symplectic and multisymplectic schemes   preserve the symplectic
structure and multisymplectic structure of the systems
respectively.

Very recently, it has been found \c{GLW01}\c{hyg1} that there
exist what is called the Euler-Lagrange cohomology in  either
classical mechanics or classical field theory and it plays very
important role for the symplectic or multisymplectic structure
preserving property in each case. It has also been studied the
difference discrete version for classical mechanics and field
theory mainly in Lagrangian formalism. For this purpose, it has
been proposed a difference discrete variational principle by
regarding the forward (or backward) difference as an entire
geometric object to deal with variation of the difference discrete
classical mechanics and field theory \c{GLW01}\c{hyg1}. In
\c{GLW02} and \c{GJLW01}\c{kw1}\c{GJLW02}, special investigation
has been made for the symplectic algorithm as well as the
symplectic and multisymplectic structure preserving in simple
element method
respectively from the cohomological point of view. In \c{GLWW01},
the multi-parameter differential approach has been introduced in
order to deal with in the same framework the variation of
functional and the exterior differential calculus in the function
space, it has been further studied the Euler-Lagrange cohomology
and its relation with symplectic and multisymplectic structure
preserving properties for classical mechanics and field theory
in both the Lagrangian and Hamiltonian formalism. The
cohomological approach has also been applied to what are called
Hamiltonian-like ODEs and PDEs respectively.

In this paper, 
we further study in some details the difference discrete
variational principle and apply it to action functional not only
in the Lagrangian formalism but also in the Hamiltonian formalism.
We also study the difference discrete versions for the
Euler-Lagrange
cohomology, symplectic and multisymplectic structures with their 
 structure preserving properties in the
classical mechanics and  field theory. We generalize the
multi-parameter differential approach for the both
variational principle   and 
exterior differential calculus in the function space to the
difference discrete variational principle and exterior
differential calculus in the function space on difference discrete
base space 
in the Lagrangian and Hamiltonian formalism. 
It is shown that the difference discrete variational principle
gives rise to the difference discrete version of the
Euler-Lagrange equations and that of the canonical equations of
motion that preserve the symplectic or multisymplectic structures
in the Lagrangian and Hamiltonian formalism for the difference
discrete mechanics and field theory, respectively. It is also
shown that the difference discrete version of the Euler-Lagrange
cohomology in each case is nontrivial and that of symplectic and
multisymplectic structures are preserved if and only if relevant
closed Euler-Lagrange conditions are satisfied without making use
of the discrete Euler-Lagrange equations or the canonical ones in
general. Although both the difference discrete Euler-Lagrange
equations or the canonical ones do satisfy the difference discrete
closed Euler-Lagrange  conditions.
Therefore, it is important that these difference discrete version
for the  symplectic and multisymplectic structure-preserving 
properties hold in the function space on the configuration space
and its tangent space in the Lagrangian formalism or on the phase
space in the Hamiltonian formalism with the relevant closed
Euler-Lagrange conditions in general rather than in the solution
space of the difference discrete Euler-Lagrange equations or that
of the canonical ones only.

 One of the key issues  of this paper
that is different from the others is the difference discrete
variational principle is first proposed in \c{GLW01}\c{hyg1} to
get difference discrete Euler-Lagrange equations. As was
emphasized in \c{GLW01}\c{hyg1}, in view of noncommutative
differential calculus, the difference is defined as the (discrete)
derivative so that it should be regarded as an entire geometric
object. Furthermore, it can also combine together in certain
manner as a geometric object to construct the numerical schemes
(see section 5). In the difference discrete variational principle
approach, this point of view has been carried out. In this paper,
this approach is applied not only to the Lagrangian formalism but
also the Hamiltonian formalism for the both difference discrete
mechanics and field theory. Together with suitable Leibniz law for
the differences, it is also directly applied to the derivation of
the numerical schemes in symplectic and multisymplectic
algorithms.

Second key issue of this paper is about the difference discrete
version of the Euler-Lagrange cohomological concepts and content
and their role-played in the symplectic and  multisymplectic
structure
preserving properties. 
in each case. 
As a matter of fact, the nontriviality of the difference discrete
version for the Euler-Lagrange cohomology plays a crucial role and
is directly related to the symplectic and multisymplectic
structure preserving properties.

In the  course of numerical calculation, the ``time"  $t \in R$ is
always discretized, say, with equal spacing $\t=\Delta t$ and the
space coordinates are also discretized in many cases, especially,
for the classical field theory. In addition to these computational
approach, there  also exist various discrete physical systems with
discrete or difference discrete Lagrangian and Hamiltonian
functional. It is well known that the differences of functions do
not obey the ordinary Leibniz law. In order to explore that the
difference discrete symplectic and multisymplectic  structures in
these difference discrete systems and their structure-preserving
properties,
 some noncommutative differential calculus  should be employed,
Even for the well-established symplectic algorithm. This is the
third key point of this paper. Recently, the noncommutative
differential calculus in regular lattice has been employed to deal
with the difference discrete phase space for finite dimensional
systems with separable Hamiltonian \c{GWWWW00}\c{GWZ00}\c{HYG99}.
Similar noncommutative differential calculus will be employed in
the present paper.

Another key point of this paper 
 is the multi-parameter differential approach to the difference
discrete variational principle and to deal with the exterior
differential calculus in the function space. This approach
provides the same framework for the both DDVP and EL cohomological
approach. It was employed in \c{GLWW01} for the continuous cases.
In the present paper we will employ the multi-parameter
differential
approach 
for the cases on the difference discrete version of the base
space, i.e. the ``time" in the difference discrete mechanics and
the ``spacetime/space" in the difference discrete field theory.

The plan of this paper is as follows. We first explore, in the
framework of the
 multi-parameter differential approach, 
the difference discrete variational principle  in the both
Lagrangian and Hamiltonian formalism for classical mechanics and
classical field theory respectively in section 2. It is shown that
difference discrete variational principle with simply modified
Leibniz law for the differences offers the difference discrete
version for both Euler-Lagrange equations and the canonical
equations of motion. In section 3 and 4,  also in the framework of
the multi-parameter differential approach and using the exterior
differential calculus in
the function space 
in the case of difference discrete base space, we deal with such
kind of the difference discrete versions of the Euler-Lagrange
cohomology as well as the symplectic structure preserving and
multisymplectic structure preserving properties in the both
Lagrangian and Hamiltonian formalism for classical mechanics and
field
theory respectively. 
 It is shown
that the relevant difference discrete versions of the
Euler-Lagrange cohomology in each case is nontrivial and it is
directly linked with the difference discrete symplectic and
multisymplectic structure preserving properties.
We explore in some details the difference discrete variational
principle approach and the difference discrete  Euler-Lagrange
cohomological approach to the symplectic and
multisymplectic algorithms in section 5. 
It is pointed out that the difference discrete the Euler-Lagrange
equations, the canonical equations of motion for the classical
mechanics and  field theory present themselves certain symplectic
and multisymplectic schemes respectively. We also show that the
Euler midpoint scheme in the symplectic algorithm, the  midpoint
box scheme for a type of PDEs and the midpoint box scheme for the
Hamiltonian field theory in the multisymplectic algorithm can be
derived by the difference discrete variational principle with a
suitable difference Leibniz law. And the difference discrete
Euler-Lagrange cohomology and its relation with the difference
discrete symplectic and multisymplectic structure preserving
properties offer a cohomological scenario to show whether
numerical schemes are symplectic or multisymplectic. Finally, we
end with some concluding remarks in section 6. In the appendix,
some simple relevant noncommutative differential calculus on
regular lattice with equal step-length on each direction are
given. For the sake of self-containing in relevant sections, the
content for the continuous case is  briefly recalled and then the
approach is generalized to deal with the difference discrete case.

\section{Variational and difference discrete variational principle in multi-parameter differential approach}

$\quad$ In order to consider certain difference discrete versions
of the  simplectic and multisymplectic structures and their
structure-preserving properties in both Lagrangian and Hamiltonian
formalism for classical mechanics and field theory, 
we study the variational principle and difference discrete
variational principle  in both Lagrangian and Hamiltonian
formalism for classical mechanics and field theory and their
difference discrete  versions, in the framework of the
multi-parameter differential approach  in this section. We
consider the cases in classical mechanics in the subsection 2.1,
and that in classical field theory in the subsection 2.2.

The difference discrete variational principle approach was first
proposed in \c{GLW01}\c{hyg1} with vanishing condition at $t_k=\pm
\infty$ for the infinitesimal variations of coordinates in the
configuration space, $\d q^i(t_k), k \in Z$, in the difference
discrete classical mechanics and the corresponding vanishing
condition at infinity in 1+1 dimensional or 2 dimensional cases
for the infinitesimal variations of a set of generic field
variables, $\d u^{\a (i,j)}, \a=1, \cdots, r, (i,j) \in Z \times
Z,$ in the difference discrete classical field theory.
As was emphasized, the most important point of the approach is
regarding the forward difference or its certain combination in
each difference discrete case as an entire geometric object in the
sense of noncommutative differential calculus. Of course, if the
backward difference is preferred rather than the forward one, the
framework is almost the same. The framework of the multi-parameter
differential approach has been employed for the continuous case in
\c{GLWW01}. We review this approach and generalize it to deal with
the difference
discrete variational principle.

\subsection{Variational  and  difference discrete variational principle in
 continuous and difference discrete classical 
 mechanics}

 $\quad$ We begin with recall some content of the multi-parameter differential
 approach to variational principle for classical Lagrangian mechanics
and transfer it to the Hamiltonian formalism. Then we
 generalize  it to deal with the difference discrete variational principle in
 the both Lagrangian and Hamiltonian formalism for
 the difference discrete classical mechanics.

\subsubsection{Variational principle in multi-parameter differential approach
 for classical 
 mechanics }

 $\quad $
 Let time $t\in R^1$ be the base manifold, $M$ the $n$-dimensional configuration space
 on $t$ with coordinates $q^i(t)$, $(i=1, \cdots, n)$,
$TM$ the tangent bundle of $M$ with coordinates $(q^i(t), \dot
q^j(t))$, where $\dot q^j(t)$ is the time derivative of $q^j$,
  $F(TM)$ the function space on $TM$.

\vskip 4mm {\parindent=0mm{\bf 2.1.1.1  
 Variational principle   in Lagrangian formalism}} \vskip 4mm

The Lagrangian of the systems is denoted by $L(q^i, {\dot q^j})$.
For simplicity, we suppose that the Lagrangian does not manifestly
depend on $t$. The action functional along a curve $q(t)$ in $M$,
${\cal C}_a^b$ with two endpoints $a$ and $b$, can be constructed
by integrating of $L$ along the tangent of the curve \be\label{S1}
S(q(t)):=\int_a^b dt L(q^i(t), {\dot q^i(t)}). \ee

Let us consider the case that at the moment $t$ both $q^i(t)$ and
${\dot q^j(t)}$ variate by an infinitesimal increments and  the
curve ${\cal C}_a^b$ becomes a congruence of curves ${{\cal
C}_{\ep a}}^b$. The infinitesimal variations of $q^i$ and ${\dot
q^j}$ in the congruence can be described as follows
\be\label{dfmq1} q^i(t) \rightarrow q_{\ep} ^i(t)=q^i(t)+{\ep^k}
\d_k q^i(t), \quad {\dot q}^j(t) \rightarrow {\dot
q_{\ep}}^j(t)={\dot q}^j(t)+{\ep^k} {\d_k {{\dot q}^j}}, \ee where
${\ep^k, k=1, \cdots, n,}$ are $n$ free parameters that each of
them corresponds one direction in the 
configuration space $M$, 
$\d_k q ^i(t)$ and $\d_k
{\dot q} ^j(t)$ infinitesimal increments
 of $q^i(t)$  and ${\dot q ^j}(t)$ at the moment $t$ along  the
 direction $k$ in the
congruence of curves ${{\cal C}_{\ep a}}^b$: \be\label{vrq1}
 \d_k q^i(t):= \f \partial {\partial \epsilon^k} \mid_{\ep^k=0} q_{\ep}^i(t), 
 \qquad
\d_k {\dot q}^j(t):=\f {\partial} {\partial \ep^k} \mid_{\ep^k=0}
{\dot
q_{\ep}}^j(t).
\ee Here the differentials of $q_{\ep} ^i(t)$ and ${\dot q_{\ep}
^j}(t)$ with respect to $\epsilon^k$
 in the function space $F(TM)$ are manipulated. Namely,
 the 
differentials of
  $q_{\ep} ^i(t)$  and ${\dot
q_{\ep} ^j}(t)$ at the moment $t$ in the congruence of curves
${{\cal C}_{\ep a}}^b$ can be calculated by: \be \label{dq1}d
q_{\ep }^i:=\f {\pa q_{\ep}^i} {\pa {\ep^l} } d{\ep^l} =d{\ep}^k
\d_k q^i,\qquad d {\dot q_{\ep }^j}:=\f {\pa \dot q_{\ep}^i} {\pa
{\ep^l} } d{\ep^l} =d{\ep}^k \d_k \dot q^i. \ee

This framework is called the
 multi-parameter differential approach. Furthermore, the exterior
 differential calculus in the function space can also be well established
in this framework. It should be mentioned that in the standard
parameter-differential approach to the variation calculation is
usually to introduce only one free parameter along the curve. For
the variation calculation it is enough, but it should have more
degree of freedom for the exterior differential
calculation for the functions and functionals. 
The multi-parameter differential setting, in fact, offers the same
framework to deal with both variation and exterior differential
calculation for the functions and functionals.

In the congruence of curves  ${{\cal C}_{\ep a}}^b$, the
Lagrangian now becomes a family of Lagrangian and the same for the
action functional: \be S(q(t)) \rightarrow S_{\ep}
(q_{\ep}(t))=\int_a^b dt L_{\ep}(q_{\ep}^i(t), {\dot
q_{\ep}^i(t)}), \ee where the upper-index $k$ of $\ep^k$ is
omitted.

Hamilton's principle,  i.e. the (least) variational principle,
seeks the curve  ${{\cal C}_{a}}^b \in  {{\cal C}_{\ep a}}^b$
along which the action $S$ is stationary against all variations of
$q^i(t)$
along any directions. 
In the multi-parameter differential approach, similar to
(\ref{vrq1}), this can be manipulated by taking differentiation
with respect to $\ep^k$  and setting ${\ep^k}=0 $ afterwards: \be
\d_k S(q(t)) :=\f {\partial} {\partial \ep^k} \mid_{\ep^k=0}
S_{\ep} (q_{\ep}(t)) =0, \quad \forall k=1, \cdots, n, \ee for all
$\d q_{\ep^k }^i(t)=\d_k {q^i(t)} $ with $\d_k q^i(a) = \d_k
q^i(b)=0$.

It is straightforward to get the differentiation  of the action
with respect to ${\ep^k}$ \be \label{dS1}
d S_{\ep} (q_{\ep}(t))=
\int_a^b dt \f {\pa} {\pa {\ep^k}}L(q_{\ep}^i(t), {\dot
q_{\ep}^i(t)}) d{\ep^k} =\int_a^b dt dq_{\ep}^i \{\frac {\pa
L_{\ep} }{\pa {q_{\ep} ^i}}-\frac {d} {dt} \frac {\pa L_{\ep} }
{\pa {\dot q_{\ep} ^i}}\}+\f {\pa L_{\ep} } {\pa  \dot q_{\ep} ^j}
dq_{\ep}^j \mid_a^b. \ee Therefore, the variation of the action
along the direction $k$ is given by \be \d_k S (t)=\int_a^bdt \d_k
q^i \{\frac {\pa L}{\pa {q^i}}-\frac {d} {dt} \frac {\pa L} {\pa
{\dot q^i}}\} +\f {\pa L} {\pa  \dot q^j} \d_k q^j\mid_a^b. \ee
The last term in the above equation vanishes due to $\d_k q^i(a) =
\d_k q^i(b)=0$, hence the stationary requirement for $S$, i.e. the
variations of $S$ along any direction should be vanish, yields the
Euler-Lagrange equations \be\label{ee1} \frac {\pa L}{\pa
{q^i}}-\frac {d} {dt} \frac {\pa L} {\pa {\dot q^i}}=0. \ee

\vskip 4mm {\parindent=0mm{\bf 2.1.1.2  
 Variational principle   in Hamiltonian formalism}} \vskip 4mm

 The
action principle can also be carried out on the phase space in the
Hamiltonian formalism. In order to transfer to the Hamiltonian
formalism, we introduce a family of conjugate momenta from the
family of Lagrangian
$L_{\epsilon}$ 
\begin{equation}\label{cmntcm}
p_{j \epsilon}= \frac {\partial L_{\epsilon}} {\partial {{\dot
{q}}^j} _{\epsilon}},
\end{equation}
and take a Legendre transformation to get the Hamiltonian  in the
family
\begin{equation}\label{ltfmcm}
H_{\epsilon}:=H({q^i}_{\epsilon}, p_{j \epsilon})=p_{k
\epsilon} {{\dot q}^k} _{\epsilon} -L({q^i}_{\epsilon}, {{\dot q}^j%
} _{\epsilon} ).
\end{equation}

Now the family of the action functionals can be expressed as
\begin{equation}\label{action functionalh1}
S_{\epsilon} =\int_{t_1}^{t_2}dt \{p_{k \epsilon} {\dot q}%
_{\epsilon}^k -H({q^i}_{\epsilon}, {p}_{j \epsilon})\}
\end{equation}The variation of the action functional along the
direction $k$ can be calculated also in terms of differentiation
with respect to the parameter $\epsilon^k$ and setting
$\epsilon^k=0$ afterwards
\begin{equation}  \label{deltaS2}
\delta_kS=\frac{\partial}{\partial\epsilon^k}S_{\epsilon}|_{\epsilon^k=0}=%
\int_{t_1}^{t_2}dt \{-(\frac {\partial H_{\epsilon} } {\partial
{p_{i
\epsilon} }}-\dot q_{\epsilon} ^i)\delta_k p_{i \epsilon} -(\frac {%
\partial H_{\epsilon} } {\partial {\ q_{\epsilon} ^i}}+\dot p_{i
\epsilon}) \delta_k q_{\epsilon} ^i+\frac{d}{dt}(p_{i
\epsilon}\delta_k q_{\epsilon} ^i)\}|_{\epsilon^k=0}.
\end{equation}
Thus, the stationary requirement for the action against all
variations along any direction, i.e. $\delta_k S=0, \forall k=1,
\cdots, n$ together with the fixed endpoint condition  lead to the
canonical equations
\begin{equation}\label{ceqs1}
\dot q ^i=\frac {\partial H } {\partial p_i }, \qquad
\dot p_i=-\frac {\partial H } {\partial q^i}.%
\end{equation}

\subsubsection{Difference discrete
variational principle in multi-parameter differential approach
  for discrete classical 
mechanics}

 $\quad$ Let us now introduce the  difference discrete variational principle \c{GLW01}\c{hyg1}  for the difference discrete version of the
 classical 
 mechanics. Here we generalize the
 multi-parameter differential approach and employ it to deal with
 the difference discrete variational principle for the difference discrete classical 
 mechanics.

 Consider the case
that ``time" $t$ is difference discretized while the
$n$-dimensional configuration space $M_k$ at each moment $t_k, k
\in Z$, is still continuous and 
smooth enough.
Let us assume, without loss generality, that 
 the ``time"  $t \in R$ be discretized as a set of nodes and links with equal
 step-length $\t=\Delta t $:
\be t\in R  \rightarrow t\in  {\cal T}=\{ (t_k , t_{k+1}=t_k+\t,
\quad k \in Z)\}. \ee

Let $\cal N$ and $\cal L$ be the set all nodes and links with
index set  $Ind({\cal N})=Ind({\cal L})=Z$, ${\cal M}=\bigcup_{k
\in Z} M_k$ the configuration space on $\cal T$ that is still
continuous and at least pierce wisely smooth enough. At the moment
$t_k$, ${\cal N}_k$ and ${\cal L}_k$ be the set of nodes and links
neighboring $t_k$ respectively. For example, ${\cal L}_k$ includes
two links
 $[t_{k-1}, t_k]$ and $[t_k, t_{k+1}]$ with endpoints $(t_{k-1}, t_k, t_{k+1})$.
 Let ${\cal I}_k $ the index set of nodes of ${\cal N}_k$ including $t_k$, ${\cal
N}_k=\bigcup_{Ind({\cal N})\in {\cal I}_k}{\cal N}$ etc. 
The coordinates of $M_k$ 
 are   denoted  by $q^i({t_k})=q^{i (k)}, i=1, \cdots,
n$.  $T(M_k)$ the tangent bundle of $M_k$ in the sense that
difference at $t_k$ is its base, $T^*(M_k^n)$ its dual.
Let ${\cal M}_k=\bigcup_{{l} \in {\cal I}_k} M_l $ be the union of
configurantion spaces $M_l$ at $t_l, {l} \in {\cal I}_k$ on ${\cal
N}_k$, $T{\cal M}_k=\bigcup_{{l}\in {\cal I}_k}TM_l $ the union of
tangent  bundles on ${\cal M}_k$, $F(TM_k)$ and $F(T{\cal M}_k)$
the function spaces on each of them respectively, etc.. In the
difference discrete variational principle, we will use these
notions. 

\vskip 4mm
 {\parindent=0mm{\bf 2.1.2.1 Difference discrete variational principle   in Lagrangian
formalism}}

\vskip 4mm
 We first study the difference discrete  version of the Lagrangian formalism. It is clear that the
difference discrete Lagrangian written as
 \be \label{LD1} {L_D}^{(k)}=L_D(q^{i (k)}, {q_{t}}^{i (k)}) \ee is a functional on
 $F(T{\cal M}_k)$, since
${q_{t}}^{i (k)}$ is the forward difference of ${q^{i (k)}}$ at
$t_k$ defined by \be\label{fd1}
 \Delta_t q^{i (k)}:=\frac {d } {dt} q^{i
(k)}={q_t}^{i (k)} =\frac 1 {\tau} \{{q^{i (k+1)}}-{q^{i (k)}}\}.
\ee It is the (discrete) derivative and the base of $T({\cal T})$
in the sense of noncommutative differential calculus on a regular
lattice $L^1$ with equal step-length $\t$ \c{GWWWW00} (see also
the appendix) and the same notation for it as in the continuous
case may be employed if it does not cause any ambiguity.

As was emphasized, in what follows the forward difference is
viewed as an entire geometric object and its dual $d_{\cal T}t$ is
the base of $T^*({\cal T})$ in the sense \be d_{\cal
T}t(\Delta_t)=1. \ee

It is well known that the (forward) difference as the discrete
derivative does not obey the Leibniz law but the modified one
\be\label{lbnz1}
 \Delta_t(f \cdot g)^{(k)}=\Delta_t
f^{(k)}\cdot g^{(k)} + f^{(k+1)}\cdot \Delta_t g^{(k)}, \quad f, g
\in FM=\Omega^0_{\cal T} .\ee On the other hand, however, it is
important to note (see the appendix) that in the  space $T^*({\cal
T})$ dual to $T({\cal T})$, an exterior differential operator
$d_{\cal T}$ exists such that \be d_{\cal T}: \Om_{\cal T} ^l
\rightarrow \Om_{\cal T} ^{l+1},\qquad d_{\cal T}^2=0, \ee where
$\Om_{\cal T}^l$ the space of $l$-forms, $l=0, 1$, on $T^*({\cal
T})$ and $d_{\cal T}$ does satisfy the Leibniz law: \be d(\om \we
\t)^{(k)}=d\om^{(k)}\we \t ^{(k)}+(-1)^{deg(\om)}\om ^{(k)} \we \t
^{(k)}. \ee

 The action
functional in the continuous case (\ref{S1}) 
now becomes \be\label{SDcm} S_D=\sum_{k \in Z}L_D(q^{i (k)}, {q_t}
^{i (k)}), \ee where the summation is taken over $k \in Z$.

 We now consider  
how to calculate the variation of the action functional $S_D$ in
this case. Since only the ``time" is discretized while either the
configuration space at each moment $t_k$, i.e. at the node $k$, or
at its neighboring union are still continuous and the variational
calculation that will be carried out is mainly local, therefore,
the difference discrete variations may still be manipulated in the
framework of the multi-parameter differential approach. In
addition, as in the continuous case of the classical mechanics,
the differential and exterior differential calculus in the
function space 
can also be carried out in either $F(TM_k)$ and $F(T{\cal M}_k)$,
etc..

In order to make use of the multi-parameter differential approach
the variations of $q^{i (k)}$ and ${q_t}^{i (k)}$ with the
multi-parameter $\epsilon^{l}$ should be introduced. At the moment
$t_k$, 
we have\be {q_{\ep }}^{i (k)}=q^{i (k)}+\ep^l\d_l q^{i (k)}, \quad
\Delta_t {q_{\ep }}^{i (k)}=\Delta_t q^{i (k)}+\ep^l\d_l (\Delta_t
q^{i (k)}),
 \ee
 and
 \be
\d_l  {q_{\ep }}^{i (k)}:=\f \partial {\partial \ep^l}|_{\ep^l=0}
{q_{\ep} }^{i (k)}=\d_l  q^{i (k)}, \quad \d_l {q_{t \ep }}^{i
(k)}:=\f \partial {\partial\ep^l}|_{\ep^l=0}{q_{t \ep }}^{i
(k)}=\d_l {q_{t }}^{i (k)}.
 \ee
Then the action functional in (\ref{SDcm}) becomes a family of
action functionals \be S_D \rightarrow S_{D \epsilon} =\sum_{k \in
Z} {L_{D \ep }}^{(k)}\ee and the variation of the action
functional along the direction $l$ \be \label{vrSd}\d_l S_{D
\epsilon}=\f \partial {\partial\ep^l}S_{D \ep
}|_{\ep^l=0},\ee%
 i.e. \be \delta_l S_{D \epsilon} =\sum_{k \in
Z}\{\frac {\pa {L_{D \ep}}^{(k)}} {\pa {q_{\ep }}^{i (k)}}\delta_l
{q_{\ep} }^{i (k)}
           +\frac {\pa {L_{D \ep}}^{(k)}} {\pa (\Delta_t {q_{\ep} }^{i
           (k)})}
           \delta_l {q_{t \ep }}^{i (k)}\}|_{\ep^l=0}.
\ee

 By virtue of the modified Leibniz law
(\ref{lbnz1}) for $\Delta_t=\pa_t$, we have \be\label{lbnz2}
\Delta_t(\frac {\pa {L_D}^{(k-1)}} {\pa (\Delta_t q^{i
(k-1)})}\d_l q^{i (k)}) = \frac {\pa {L_D}^{(k)}} {\pa q^{i
(k)}}\d_l {q_t}^{i (k)}
 +\Delta_t(\frac {\pa {L_D}^{(k-1)}} {\pa (\Delta_t q^{i (k-1)})})\d_l q^{i (k)}.
\ee Therefore,
$$
\d_l S_{D}=\sum_{k \in Z}\{(\frac {\pa {L_{D \ep}}^{(k)}} {\pa
{q_{\ep}}^{i (k)}}
      -\Delta_t(\frac {\pa {L_{D \ep}}^{(k-1)}} {\pa (\Delta_t {q_{\ep}}^{i (k-1)})}))
      \d_l {q_{\ep}}^{i (k)}\}|_{\ep^l=0}
+\sum_{k \in Z}
 \Delta_t(\frac {\pa {L_{D \ep}}^{(k-1)}} {\pa ({{\Delta_t q}_{\ep}}^{i (k-1)})}
 \d_l {q_{\ep}}^{i (k)})|_{\ep^l=0}.
$$
Using the properties (see the appendix) \be \sum_{k \in Z}\Delta_t
f(t_k)=f(t_{k=+\infty})-f(t_{k=-\infty}),
\ee and assuming $\d_l q^{i (k)}|_{k\pm\infty}=0$, 
it follows the discrete Euler-Lagrange  equations \be \label{del1}
\frac {\pa {L_D}^{(k)}} {\pa q^{i (k)}}
           -\Delta_t(\frac {\pa {L_D}^{(k-1)}} {\pa ({\Delta_t q^{i (k-1)}})})=0.
\ee

It should be mentioned here in general, for the forward difference
calculation more general Leibniz law can be adopted and it will
lead to more general difference discrete  version of the
Euler-Lagrange equations. We will explore this issue mainly in the
Hamiltonian formalism in the
section 5.

Let us consider an example.

{\it Example 2.1. A difference discrete classical mechanics}

Consider a difference discrete version of classical mechanics with
following difference discrete Lagrangian: \be\label{lexmple1}
{L_D}
(q^{i (k)}, {\Delta_t q}^{j (k)})
 =\frac 1 2
({\Delta_t q}^{i (k)})^2-V(q^{i (k)}). \ee The difference discrete
variational principle   gives the discrete Euler-Lagrange equation
\be \label{del2} \Delta_t(\Delta_t q^{i (k-1)})-\f {\pa} {\pa q}
V(q^{i (k)})=0, \ee i.e. \be \frac 1 {\t ^2} (q^{i (k+1)}-2q^{i
(k)}+q^{i (k-1)}) =\f {\pa} {\pa q} V(q^{i (k)}). \ee This is the
difference discrete counterpart of the equation in the continuous
case. It has correct continuous limit.

\vskip 4mm
{\parindent=0mm{\bf 2.1.2.2  Difference discrete variational
principle in Hamiltonian formalism}}

\vskip 4mm
Now we consider the difference discrete variational principle on
the phase space in the difference discrete (``time") Hamiltonian
formalism.

To transfer to the difference discrete  Hamiltonian formalism, we
first define a family of the discrete canonical conjugate momenta
\be {p_{i \epsilon }}^{(k)}=\frac {\pa {L_{D \epsilon}}^{(k-1)}}
{\pa ({\Delta_t q_{\epsilon}}^{i (k-1)})}. \ee Then a family of
the difference discrete Hamiltonian can be introduced through the
discrete Legendre transformation in the family \be\label{dltn1}
{H_{D \epsilon}}^{(k)}={p_{i \epsilon }}^{(k+1)}\Delta_t
q_{\epsilon}^{i (k)}-{L_{D \epsilon}} ^{(k)}. \ee

Now the difference discrete  version of the action functional in
(\ref{SDcm}) becomes a family of action functionals: 
\be\label{SDhm} S_D \rightarrow S_{D \epsilon}=\sum_{k \in
Z}\{{p_{i \epsilon }}^{(k+1)}\Delta_t q_{\epsilon}^{i (k)}-{H_{D
\epsilon}} ^{(k)}\}.\ee Then the variation of the action  along
the direction $l$ in (\ref{SDhm}) can be calculated as \be
\delta_l S=\frac{\partial}{\partial\epsilon^l}S_{D
\epsilon}|_{\epsilon^l=0}.\ee For the differential of $S_{D
\epsilon}$ in the above equation, we have \be d S_{D
\epsilon}=\sum_{k \in Z}\{{d p_{i \epsilon }}^{(k+1)}(\Delta_t
q_{\epsilon}^{i (k)}-\frac{\partial {H_{D \epsilon}}
^{(k)}}{\partial  p_{i \epsilon }^{(k+1)} })-(\Delta_t p_{i
\epsilon }^{(k)}+\frac{\partial {H_{D \epsilon}} ^{(k)}}{\partial
q_{\epsilon}^{i (k)}}) d q_{\epsilon}^{i (k)}+ \Delta_t (p_{i
\epsilon }^{(k)}d q_{\epsilon}^{i (k)})\}.\ee Here, the modified
Leibniz law (\ref{lbnz1}) has been used.

Now, the difference discrete variational principle gives rise to
the difference discrete  version of the canonical equations of
motion (\ref{ceqs1})\be\label{ceqsd1}\Delta_t q^{i
(k)}=\frac{\partial {H_D } ^{(k)}}{\partial {p_i }^{(k+1)}
},\qquad \Delta_t {p_i }^{(k)}=-\frac {\partial
{H_D}^{(k)}}{\partial q^{i (k)}}.\ee

In fact, the first set of equations above can directly be derived
from the Legendge transformation (\ref{dltn1}) and the second set
can be gotten from the Legendge transformation (\ref{dltn1}) and
the Euler-Lagrange equations (\ref{del1}). This indicates that the
difference discrete variational principle approach to the
difference discrete version of classical mechanics is
self-consistent.

Let us consider the example 2.1 in the discrete Hamiltonian
formalism.

{\it Example 2.2. Hamiltonian formalism for the example 2.1. }

 First, the difference discrete conjugate momentum is introduced
\be
 {p_i}^{(k)}=\frac {\pa {L_D}^{(k-1)}} {\pa ({\n_t q}^{i (k-1)})}=\Delta_t q^{i (k-1)}.
\ee The Hamiltonian is introduced through the discrete Legendre
transformation \be {H_D}^{(k)}={p_i}^{(k+1)}{\Delta_t q^{i
(k)}}-{L_D}^{(k)} =\f 1 2 {{p_j}^{(k+1)}}^2+V(q^{i (k)}). \ee And
a pair of difference discrete canonical equations read now \be
\label{dHE1}\Delta_t q^{i (k)}={{p_i}^{(k+1)}},\qquad \Delta_t
{p_i}^{(k)}=-\f {\pa} {\pa q} V(q^{i (k)}) \ee In fact, the  time
difference discrete derivative of ${p_i}^{(k)}$ can also be
derived from the difference discrete Lagrangian and the discrete
Euler-Lagrange equation (\ref{del2}).

\subsection{Variational and difference discrete variational principle in
 continuous and difference discrete classical 
 field theory}

$\quad $ We now study the difference discrete variational
principle
 in Lagrangian and Hamiltonian
formalism for the difference discrete classical field theory. 
For the sake of simplicity, let us consider  the 1+1-d and 2-d
cases in discrete  classical field theory for  a set of generic
fields $u^{\a}, \a=1, \cdots, r$. We first recall the
multi-parameter differential approach to the variation of
functional in  Lagrangian  formalism \c{GLWW01} and deal with the
Hamiltonian formalism, then generalize it to  the difference
discrete variational principle for the difference discrete
classical field theory  in both Lagrangian and Hamiltonian
formalism.

\subsubsection{Variational principle in multi-parameter differential approach
for classical 
field theory}

$\quad $
For the sake of simplicity, let 
$X^{(1,n-1)}$ be an $n$-dimensional Minkowskian space as base
manifold with coordinates $x^{\mu}$, $(\mu =0, \cdots,n-1 )$, $M$
the configuration space
 on $X^{(1,n-1)}$ with a set of generic fields $u^{\a}(x)$, $(\a=1, \cdots, r)$,
$TM$ the tangent bundle of $M$ with coordinates $(u^{\a},
u_{\mu}^{\a})$, where
 $u_{\mu}^{\a}=\frac {\pa u^{\a}} {\pa x^{\mu}}$,
  $F(TM)$ the function space on $TM$ etc. We also assume
these fields to be  free of constraints. In fact, the approach
here can easily be applied to other cases.

\vskip 4mm
{\parindent=0mm{\bf 2.2.1.1  Variational principle in Lagrangian
formalism}}

\vskip 4mm

The Lagrangian of the fields now is a functional of the set of
generic fields under consideration: \be L(u^{\a}, \dot
u^{\a})=\int d^{n-1}x {\cal L}(u^{\a}({\bf x},t),
u_{\mu}^{\a}({\bf x}, t)),\qquad u^{\a}(x)=u^{\a}({\bf x},t),
\quad etc., \ee and the action is given by \be S(u^{\a})
=\int dt L( u^{\a}, \dot u^{\a})=\int d^nx {\cal
L}(u^{\a}(x), u_{\mu}^{\a}(x)), \ee where ${\cal L}(u^{\a}(x),
u_{\mu}^{\a}( x))$ is the Lagrangian density.

In order to apply Hamilton's principle we first consider how to
define the variation of the action functional $S(u^{\a})
$ in a manner analog to the case of classical mechanics. In order
to achieve this purpose, let us suppose that both $u^{\a}(x)$ and
${u_{\mu}^{\a}(x)}$ variate by an infinitesimal increments such
that at a spacetime point of $x$ the infinitesimal variations of
$u^{\a}$ and ${u_{\mu}^{\a}}$ can be described as follows \be
u_{\ep}^{\a}(x)=u^{\a}(x)+{\ep^{\b}} \d_{\b} u^{\a}(x), \qquad
{u_{\mu \ep}^{\a}}(x)=u_{\mu}^{\a}(x)+{\ep^{\b}}{\d_{\b} {u_{\mu }
^{\a}}(x)}, \ee where ${\ep^{\b}}, {\b}=1, \cdots, r$ are free
parameters, 
each of which corresponds one direction in 
 the configuration space $M$, and  \be \d_\beta
u_{\ep}^{\a}(x):=\f
\partial {\partial \ep^{\b}}\mid_{\ep^{\b}=0} u_{\ep}^{\a}(x),
\qquad \d_\beta u_{\mu \ep}^{\a}(x):=\f \partial {\partial
\ep^{\b}}\mid_{\ep^{\b}=0}u_{\mu \ep}^{\a}(x),
\ee the infinitesimal increments
 of $u^{\a}(x)$  and $u_{\mu}^{\a}(x)$ along the direction ${\b}$ at the
spacetime point $x$
 respectively.

 On the other hand, similar to the case of classical mechanics, the differential
 calculus of  $u^{\a}(x)$ and $u_{\mu}^{\a}(x)$
 in the function space $F(TM)$ can also be manipulated in the framework of the
 multi-parameter differential approach. Furthermore, the exterior
 differential calculus in this framework 
can also be well established.

 Thus the (exterior) differentials of
 $u^{\a}(x)$  and $u_{\mu}^{\a}(x)$ in the function space $F(TM)$ at the spacetime point
 $x$ can be defined as:
\be
 du_{\ep}^{\a}(x):= \f {\pa u_{\ep}^{\a}(x)} {\pa {\ep^{\b}} }  d{\ep^{\b}}
=d{\ep}^{\b} \d_{\b} u^{\a},\qquad d{u_{\mu \ep}^{\a}(x)}:= \f
{\pa u_{\mu \ep}^{\a}(x)} {\pa {\ep^{\b}} }d{\ep^{\b}}
 =d{\ep}^{\b} \d_{\b} u_{\mu}^{\a}.%
\ee

Now, the Lagrangian also becomes a family of Lagrangian
functionals \be\label{Lft} L_{\ep}(u_{\ep}^{\a}, \dot u_{\ep}^{\a}
)= \int d^{n-1}x {\cal L}(u_{\ep}^{\a}({\bf x},t), {u_{\mu
\ep}^{\a}({\bf x},t)}), \ee and the action $S(u^{\a})$ 
 becomes a family of functionals as well \be S
\rightarrow S_{\ep}=S(u_{\ep}^{\a})
. \ee
Then the variation of the action along the direction ${\b}$ can be
manipulated as the derivative of $S_{\ep}$ with respect to
$\ep^{\b}$ and setting $\ep^{\b}=0$ afterwards. Namely,
\be\label{voa2} \d_{\b} S:=\f \partial {\partial
\ep^{\b}}\mid_{\ep^{\b}=0} S_{\ep}. \ee

Manipulating the variation of the action functional in this manner
and integrating by parts, it  follows that \be \d_{\b} S =\int
d^nx\{(\f {\pa \cal L_{\ep}} {\pa u_{\ep}^{\a}} -\pa_{\mu}({ \f
{\pa \cal L_{\ep}} {\pa u_{\mu \ep}^{\a}}}))\d_{\b} u_{\ep}^{\a}
+\pa_{\mu}(\f {\pa \cal L} {\pa u_{\mu \ep}^{\a}} \d_{\b}
u_{\ep}^{\a})\}\mid_{\ep^{\b}=0}. \ee Assuming $ \d_{\b}
u_{\ep}^{\a}\mid_{\pm \infty}=0$, and requiring $ \d_{\b} S=0$
along all directions according to Hamilton's principle, then the
Euler-Lagrange equations follow: \be \label{elenft} \f {\pa \cal
L} {\pa u^{\a}} -\pa_{\mu}({ \f {\pa \cal L} {\pa
u_{\mu}^{\a}}})=0. \ee

\vskip 4mm
{\parindent=0mm{\bf 2.2.1.2  Variational principle in Hamiltonian
formalism}}

\vskip 4mm In order to use the multi-parameter differential
approach for the variational principle in Hamiltonian formalism
for the classical field theory, we first define a family of a set
of ``momenta" that are canonically conjugate to the family of
field variables \be \pi_{\b \epsilon }(x)=\frac {\pa {\cal
L_{\epsilon}}} {\pa \dot {u}^{\b}_{\epsilon}}, \ee and take a
Legendre transformation to get the Hamiltonian density in the
family \be \label{ltfmft}{\cal H}_{\epsilon}(u^{\a}_{\epsilon},
\pi_{\a \epsilon}, {\nabla_a u}^{\a}_{\epsilon})=\pi_{\a
\epsilon}(x) {\dot u}^{\a}_{\epsilon}(x) - {\cal
L}_{\epsilon}(u^{\a}_{\epsilon},
{\dot u}^{\a}_{\epsilon}, {\nabla_a u}^{\a}_{\epsilon}), 
\ee where $\nabla_a =\frac{\partial}{\partial x^a}, a=1, \cdots,
n-1$. A family of the Hamiltonian then is given by \be
H_{\epsilon}(t)=\int d^{n-1}x {\cal H}_{\epsilon}(x), \ee with the
Legendre transformation \be H_{\epsilon}(t)=\int d^{n-1}x\pi_{\a
\epsilon}(x) {\dot u}^{\a}_{\epsilon}(x) -L_{\epsilon}(t). \ee

The action $S(u^{\a})$ 
 becomes a family of
functionals as well \be S \rightarrow S_{\ep}
=\int d^n x\{\pi_{\a \epsilon}(x) {\dot
u}^{\a}_{\epsilon}(x) - {\cal H}_{\epsilon}(u^{\a}_{\epsilon},
{\dot u}^{\a}_{\epsilon},{\nabla_a u}^{\a}_{\epsilon})\}. \ee Then
the variation of the action along the direction $\b$ can be
manipulated as the derivative of $S_{\ep}$ with respect to
$\ep^{\b}$ and setting $\ep^{\b}=0$ afterwards as was shown in
(\ref{voa2}). Namely, \be\label{dShft}\begin{array}{l}
\delta_{\beta} S=
\int d^n x\{\frac{\partial}{\partial\epsilon^{\beta}}\pi_{\a
\epsilon} ({\dot u}^{\a}_{\epsilon} -
 \frac{\partial {\cal H}_{\epsilon}}{\partial \pi_{\a
\epsilon}})-\frac{\partial}{\partial\epsilon^{\beta}}u^{\a}_{\epsilon}(x)({\dot
\pi}_{\a \epsilon}+\frac{\partial {\cal H}_{\epsilon}}{\partial
u^{\a}_{\epsilon} }-\nabla_a (\frac{\partial{\cal
H}_{\epsilon}}{\partial(\nabla_a
u^{\a}_{\epsilon})}))\\[4mm]
\qquad \quad +{\frac{\partial}{\partial t}}{(\pi_{\a
\epsilon}\frac{\partial}{\partial\epsilon^{\beta}}u^{\a}_{\epsilon})}-\nabla_a(\frac{\partial{\cal
H}_{\epsilon}}{\partial(\nabla_a
u^{\a}_{\epsilon})}\frac{\partial}{\partial\epsilon^{\beta}}u^{\a}_{\epsilon})
\}|_{\epsilon^{\b}=0}
. \end{array}\ee Thus, the canonical equations of motion follow
form the stationary requirement of the action principle \be
\label{ceqsft}{\dot u}^{\a}(x)=
 \frac{\partial {\cal H}}{\partial \pi_{\a
}}, \qquad {\dot \pi}_{\a }(x)=-\frac{\partial {\cal H}}{\partial
u^{\a} }+\nabla_a (\frac {\partial{\cal H}}{\partial(\nabla_a
u^{\a})}).\ee

\subsubsection{Difference discrete  variational principle in multi-parameter differential approach
  for discrete classical  field theory}

$\quad$ For the sake of simplicity, we consider the case of
1+1-dimensional spacetime or 2-dimensional space. It is
straightforward to generalize for higher dimensional case.

 Let
$X^{(1,1)}$ or $X^{(2)}$ with suitable signature of the metrics be
the base manifold, $L^2={\cal X}$ a regular lattice with
2-directions $x_{\mu},(\mu=1, 2)$  on $X^{(1,1)}$ or $X^{(2)}$,
$\cal N$ the all nodes on $L^2$ that are coordinated by
$x_{(i,j)}, (i,j) \in {Z \times Z}$ with index set $Ind({\cal
N})$,
$M_D:=M_{(i,j)}$ the pierce of  
configuration space with a set of generic field variables
$u^{\a}(x_{(i,j)})=u^{\a (i,j)} \in M_D$ at the node $x_{(i,j)}$,
$T{M_{(i,j)}}$ the 
tangent bundle of $M_{(i,j)}$ with the set of field variables and
their differences $(u^{\a (i,j)}, u_{\mu}^{\a (i,j)})  \in
T(M_{(i,j)})$,
$F(T{M_{(i,j)}})$ the function space on $T{M_{(i,j)}}$, etc..

For a given node with coordinates
$x_{(i,j)}$, let 
 ${\cal N}_{(i,j)}$ be the set of nodes neighboring to
$x_{(i,j)}$ with index set ${\cal I}_{(i,j)}=Ind({\cal
N})_{(i,j)}$, ${\cal X}_{(i,j)}=\bigcup_{Ind({\cal N}) \in {{\cal
I}_{(i,j)}}}{\cal N}$ a set of nodes that is related to
$x_{(i,j)}$ by the differences, ${\cal M}_D:={\cal M}_{{\cal
X}_{(i,j)}}=\bigcup_{Ind({\cal N}) \in {\cal I}_{(i,j)}}{{\cal
M}_{\cal N}}$ the union of the pierces of configuration space on
${\cal X}_{(i,j)}$.

It is known that the forward differences along each direction in
$F(T{{\cal M}_{{\cal X}_{(i,j)}}})$ are defined by
$\Delta_{\mu}u^{(i,j)}=u_{\mu}^{(i,j)}$: \be \Delta_1 u^{(i,j)}
=\frac 1 {h_1} (u^{(i+1,j)} -u^{(i,j)} ),\quad \Delta_2 u^{(i,j)}
=\frac 1 {h_2} (u^{(i,j+1)} -u^{(i,j)} ). \ee They are the bases
of $T({\cal X})$ and the upper-indexes reflect the corresponding
coordinates of nodes on ${\cal X}$. And their dual $d
x^{\mu}=d_{\cal X} x^{\mu}$ are the bases of $T^*({\cal X})$ \be
d_{\cal X} x^{\mu}(\Delta_{\nu})=\d_{\nu}^{\mu}. \ee

As in the previous subsection, the (forward) differences as the
discrete derivatives do not obey the Leibniz law but the modified
one (\ref{lbnz1}) along each direction.
While in the  space $T^*({\cal X})$ dual to $T({\cal X})$,
an exterior differential calculus can be introduced (see the
appendix) such that there exists an operator $d_{\cal X}$ with the
following properties \be d_{\cal X}: \Om ^{l (i,j)} \rightarrow
\Om ^{{l+1} (i,j)},\qquad d_{\cal X}^2=0, \ee where $\Om ^{l
(i,j)}$ is the space of all $l$-forms in $T^*({\cal X})$ and
$d_{\cal X}$ does satisfy the Leibniz law: \be d_{\cal X}(\om \we
\t)^{(i,j)}=d_{\cal X}\om^{(i,j)}\we \t
^{(i,j)}+(-1)^{deg(\om)}\om ^{(i,j)} \we d_{\cal X}\t ^{(i,j)}.
\ee

It is important to note that although the base manifold is
discretized either the configuration space at each node or its
neighboring union is still continuous. In addition, the
variational calculation that will be carried out is mainly local.
Therefore, similar to the case of difference discrete classical
mechanics, in what follows the difference discrete variations will
be manipulated on the framework of the multi-parameter
differential approach.

In addition, as was pointed out in the difference discrete version
of classical mechanics,  the differential and the exterior
differential calculus
in the function space 
can also be carried out in either $F(T{M_{(i,j)}})$ or $F(T{{\cal
M}_{{\cal X}_{(i,j)}}}):=\bigcup_{{\cal N}\in {\cal I} }F(T{\cal
N})$, etc.  on the framework of the multi-parameter differential
approach.

\vskip 4mm
{\parindent=0mm{\bf 2.2.2.1  Difference discrete variational
principle in Lagrangian formalism}}

\vskip 4mm

For the difference discrete  version of the classical field
theory, the difference discrete Lagrangian denoted as \be {{\cal
L}_D}^{(i,j)}={\cal L}_D( u^{\a (i,j)} ,u_{\mu}^{\a (i,j)} ) \ee
is a functional in $F(T{{\cal M}_{{\cal X}_{(i,j)}}})$. The action
functional is given by \be \label{action
functionalld}S_D=\sum_{(i,j) \in {Z \times Z}}{\cal L}_D(u^{\a
(i,j)}, u_{\mu}^{\a (i,j)}). \ee

Taking the variation of $S_D$ by the multi-parameter
differential approach, 
the variation along the direction $\beta$ is given by
\be\label{voald}\delta_{\beta}S_{D
}=\frac{\partial}{\partial\epsilon^{\beta}}S_{D
\epsilon}|_{\epsilon^{\beta}=0}. \ee For the sake of simplicity,
in what follows we omit the multi-parameters $\epsilon^{\beta}$ in
the course of calculation. Thus, we have
$$ \delta_{\beta} S_D =\sum_{(i,j) \in {Z \times Z}} \{\frac
{\pa {\cal L}_D^{(i,j)}} {\pa u^{\a (i,j)}} \d_{\beta} u^{\a
(i,j)}
       +\frac {\pa {\cal
L}_D^{(i,j)}} {\pa u_{\mu}^{\a (i,j)}} \d_{\beta} u_{\mu}^{\a
(i,j)}\}.
$$
Employing the modified Leibniz law (\ref{lbnz1}) for the forward
difference, we have
$$
\Delta_1(\frac {\pa {\cal L}_D^{(i-1,j)}} {\pa u_{1}^{\a (k-1,l)}}
\d_{\beta} u^{\a (k,l)}) =\frac {\pa {\cal L}_D^{(i,j)}} {\pa
u_{1}^{\a (k,l)}} \d_{\beta} u_{1}^{\a (k,l)} +\Delta_1 (\frac
{\pa {\cal L}_D^{(i-1,j)}} {\pa u_{1}^{\a (k-1,l)}}) \d_{\beta}
u^{\a (k,l)},$$$$ \Delta_2(\frac {\pa {\cal L}_D^{(i,j-1)}} {\pa
u_{2}^{\a (k,l-1)}} \d_{\beta} u^{\a (k,l)}) =\frac {\pa {\cal
L}_D^{(i,j)}} {\pa u_{2}^{\a (k,l)}} \d_{\beta} u_{2}^{\a (k,l)}
+\Delta_2(\frac {\pa {\cal L}_D^{(i,j-1)}} {\pa u_{2}^{\a
(k,l-1)}}) \d_{\beta} u^{\a (k,l)}.
$$
Assuming that $\d_{\beta} u^{\a (k,l)}$'s vanish at infinity, it
follows the discrete Euler-Lagrange equations \be \label{del3}
\frac {\pa {{\cal L}_D}^{(i,j)}} {\pa u^{\a (k,l)}} -\Delta_1
(\frac {\pa {{\cal L}_D}^{(i-1,j)}} {\pa (\Delta_1 u^{\a
(k-1,l)})}) -\Delta_2(\frac {\pa {{\cal L}_D}^{(i,j-1)}} {\pa
(\Delta_2 u^{\a (k,l-1)})}) =0. \ee

Let us consider an example to show that the difference discrete
variational principle   gives right result.

{\it Example 2.3. A discrete  classical field theory with
difference discrete Lagrangian}

Consider the following discrete  classical field theory with
difference discrete Lagrangian: \be {\cal
L}_D
( u^{\a (i,j)} ,u_{\mu}^{\a (i,j)} ) =\frac 1 2 (\Delta_{\mu}
u^{\a (i,j)})^2-V( u^{\a (i,j)}). \ee The difference discrete
variational principle   gives the discrete Euler-Lagrange equation
(\ref{del3}) as follows \be \label{del4} \Delta_1(\Delta_1 u^{\a
(i-1,j)})+\Delta_2(\Delta_2  u^{\a (i,j-1)}) -\f {\pa} {\pa
{u^{\a}}} V( u^{\a (i,j)})=0, \ee i.e. \be \frac 1 {h_1^2} (u^{\a
(i+1,j)}-2u^{\a (i,j)}+u^{\a (i-1,j)}) +\frac 1 {h_2^2} (u^{\a
(i,j+1)}-2u^{\a (i,j)}+u^{\a (i,j-1)}) =\f {\pa} {\pa {u^{\a}}}
V(u^{\a (i,j)}). \ee This is also what is wanted for the
difference discrete counterpart of the relevant Euler-Lagrange
equation in the continuous limit.

\vskip 4mm
{\parindent=0mm{\bf 2.2.2.2  Difference discrete variational
principle in Hamiltonian formalism}}

\vskip 4mm Let $X^{(1,1)}$ be the base space. We first define a
set of the discrete canonical conjugate momenta on the tangent
space of the set of nodes neighboring to the node $x(i,j)$, i.e.
on $T({\cal X}_{(i,j)})$:
 \be {\pi_{\a}}^{
(i,j)}=\frac {\pa {{\cal L}_D}^{(i-1,j)}} {\pa ({\Delta_t {u}^{\a
(i-1,j)})}}. \ee
 The difference discrete Hamiltonian is introduced through
the discrete Legendre transformation \be\label{ltfmfthd} {{\cal H}_D}^{(i,j)}
={\pi_{\a}}^{(i+1,j)}{\Delta_t {u}^{\a (i,j)}}-{{\cal
L}_D}^{(i,j)}. \ee The action functional (\ref{action
functionalld}) now is given by \be\label{action functionalhd}
S_D=\sum_{(i,j) \in {Z \times Z}}({\pi_{\a}}^{(i+1,j)}{\Delta_t
{u}^{\a (i,j)}}-{{\cal H}_D}^{(i,j)}).\ee

Taking the variation of $S_D$ by the multi-parameter differential
approach 
and using the modified Leibniz law, the variation along the
direction $\beta$ is given by \be\label{voahd}\begin{array}{l}
\delta_{\beta} S_D =\sum_{(i,j) \in {Z \times Z}}
\{\delta_{\beta}{\pi_{\a}}^{(i+1,j)}({\Delta_t {u}^{\a
(i,j)}}-\frac{\partial{{\cal H}_D}^{(i,j)}}{\partial
{\pi_{\a}}^{(i+1,j)}})-({\Delta_t \pi_{\a}}^{(i,j)} +\frac {\pa
{\cal
H }_D^{(i,j)}} {\pa u^{\a (i,j)}}\\[4mm]-\Delta_x(\frac {\pa {\cal H
}_D^{(i,j-1)}} {\pa (\Delta_x u^{\a (i,j-1)})})) \d_{\beta} u^{\a
(i,j)}
    +\Delta_t({\pi_{\a}}^{(i,j)}\d_{\beta} u^{\a
(i,j)})-\Delta_x(\frac {\pa {\cal H}_D^{(i,j-1)}} {\pa (\Delta_x
u^{\a (i,j-1)})}
        \d_{\beta} u^{\a (i,j)})\}.\end{array}
\ee The stationary requirement for difference discrete  version of
action functional against all variations along any direction gives
rise to the difference discrete version of the canonical equations
of motion: \be\label{ceqsfd}\begin{array}{l} {\Delta_t {u}^{\a
(i,j)}}=\frac{\partial{{\cal
H}_D}^{(i,j)}}{\partial {\pi_{\a}}^{(i+1,j)}},\\[4mm]{\Delta_t
\pi_{\a}}^{(i,j)}=-\frac {\pa {\cal H }_D^{(i,j)}} {\pa u^{\a
(i,j)}}+\Delta_x(\frac {\pa {\cal H }_D^{(i,j-1)}} {\pa (\Delta_x
u^{\a (i,j-1)})}). \end{array}\ee

It should be noted that the first set of equations above may
directly follow from the difference discrete  Legendre
transformation (\ref{ltfmfthd}), while the second set from the
transformation (\ref{ltfmfthd}) and the difference discrete
Euler-Lagrange equations (\ref{del3}). As in the case of
difference discrete version of the classical mechanics, this also
indicates that the difference discrete variational principle
approach is self-consistent.

 Let us consider the
example 2.3 in the discrete Hamiltonian formalism.

{\it Example 2.4: Difference discrete Hamiltonian formalism for
the example 2.3.}

 First, a set of the difference discrete conjugate momenta are introduced
\be {\pi_{\a}}^{(i,j)}=\frac {\pa {{\cal L}_D}^{(i-1,j)}} {\pa
{u}^{\a (i,j)}}=\Delta_t {u}^{\a (i-1,j)}. \ee Then the
Hamiltonian is introduced through the discrete Legendre
transformation \be {{\cal
H}_D}^{(i,j)}={\pi_{\a}}^{(i+1,j)}{\Delta_t {u}^{\a (i,j)}}-{{\cal
L}_D}^{(i,j)} =\f 1 2 {{\pi_{\a}}^{(i+1,j)}}^2+V(u^{\a (i,j)}).
\ee And a pair sets of difference discrete canonical equations
read now \be \Delta_t
{u}^{\a (i,j)}=
{\pi}^{\a (i+1,j)}, \qquad \Delta_t {\pi_{\a}}^{(i,j)}=-\f {\pa
V({u}^{(i,j)})} {\pa u^{\a (i,j)}}. \ee In fact, the  time
discrete derivative of ${\pi_{\a}}^{(i,j)}$ follows from the
difference discrete Lagrangian and the discrete Euler-Lagrange
equations (\ref{del4}).

\section{Euler-Lagrange cohomology,  symplectic structure preserving property in
continuous and discrete classical mechanics}

$\quad$ Now we are ready to study certain difference discrete
version of the Euler-Lagrange cohomology, its relations to the
simplectic and multisymplectic structures as well as their
preserving properties in the classical mechanics and field theory
 in both Lagrangian and Hamiltonian formalism.
In this  and the next section, we deal with the case of classical
mechanics and field theory respectively. We first recall some
content on the issues in the continuous case for the mechanics in
subsection 3.1. Then we consider these issues in difference
discrete classical mechanics in the Lagrangian and Hamiltonian
formalism in subsection 3.2 and 3.3 respectively.

\subsection{Euler-Lagrange cohomology, symplectic structure preserving property in
continuous classical mechanics}

$\quad$ By virtue of the multi-parameter differential approach to
exterior derivatives in the function space,  the Euler-Lagrange
cohomology and symplectic structure preserving property in
Lagrangian formalism for the classical mechanics has been studied
in \c{GLWW01}. We recall the content  in 3.1.1 and deal with the
the relevant issues in the Hamiltonian formalism in 3.1.2.

\subsubsection{
 Lagrangian formalism} 

 $\quad$ It is obvious but important
to note that in   manipulating the variation of the action
functional
 for the classical mechanics in Lagrangian
formalism in the subsection 2.1, the variation of the action is
very closely linked with the differential of $S_{\ep} $ with
respect to the free parameters $\ep^k$ in the congruence of
curves, i.e. $dS_{\ep} $.

Furthermore,  the exterior
 differential calculus in the framework of the
 multi-parameter space can  be well established. Namely, in the free
 multi-parameter space, the standard exterior differential calculus
can be introduced. 
And this multi-parameter exterior differential calculus can also
be employed to deal with the exterior differential calculus in the
function spaces on $M$ and $TM$, i.e. $FM$ and $F(TM)$. 

Therefore, the integrand in ({\ref{dS1}}) for $d S_{\ep} $ and the
boundary term may be views as 1-forms. 
And the differential 
of the family of Lagrangian functions $
L_{\ep}:=L(q_{\ep}^i(t), {\dot q_{\ep}^i(t)})$ with respect to
$\ep^k$ is given by 
\be
dL_{\ep}\mid_{\ep^k=0} =\{\frac {\pa L}{\pa {q^i}}-\frac {d} {dt}
\frac {\pa L} {\pa {\dot q^i}}\} dq_{\ep}^i +\frac {d} {dt}
\{\frac {\pa L} {\pa {\dot q^i} } dq_{\ep}^i\}. \ee

Let us 
define a family of the Euler-Lagrange  1-forms and that of the
canonical 1-forms
 $\theta$ on $T^*M$,
\be \label{E1} E_{\ep}(q_{\ep}^i, {\dot q}_{\ep}^i):=\{\frac {\pa
L_{\ep}}{\pa {q_{\ep}^i}}-\frac {d} {dt} \frac {\pa L_{\ep}} {\pa
{{\dot q}_{\ep}^i}}\} dq_{\ep}^i, \ee \be \label {t1} \theta_{L
\ep} := \frac {\pa L_{\ep}} {\pa {{\dot q}_{\ep}^i} } dq_{\ep}^i,
\ee we have \be \label{dL1} dL_{\ep}(q_{\ep}^i, \dot
q_{\ep}^i)\mid_{\ep^k=0}=E(q^i, \dot q^i) +\frac {d} {dt}
\theta_L. \ee

Furthermore, owing to the nilpotency of $d$ with respect to
$\ep^k$ in the cotangent space of the congruence of curves on
$F(T^*M)$,
$$
d^2L_{\ep}(q_{\ep}^i, {\dot q_{\ep}^j})\mid_{\ep^k=0}=0,
$$
it follows that
 \be\label{ddLcml}
dE(q^i, \dot q^i) +\frac {d} {dt} \om_L =0, \ee where  $\om_L$ is
the symplectic 2-form in the Lagrangian formalism defined by \be
\omega_L = d\theta_L = \frac {\pa^2 L} {{\pa q^j} \pa {\dot q^i} }
dq^j \we dq^i +\frac {\pa^2 L} {\pa {\dot q^j}{\pa {\dot q^i}}}
d{\dot q}^j \we dq^i . \ee And it does not change if the canonical
1-form transforms as \be {\theta} \rightarrow
{\theta_L}'={\theta_L}+d\b(q^i, \dot q^i), \ee where $\b(q^i, \dot
q^i)$ is an arbitrary function of $(q^i, \dot q^i)$.

We have established the important and significant issues on the
Euler-Lagrange cohomology and symplectic structure preserving law
in the Lagrangian formalism. Before enumerating them and exploring
their significance, let us investigate the relevant issues in the
Hamiltonian formalism.

\subsubsection{
Hamiltonian mechanics}

$\quad $ The above  cohomological and other issues 
in the Lagrangian
formalism of the classical mechanics can also be well established
 in the Hamiltonian formalism. Again the multi-parameter
 differential approach will be employed.

In order to transfer to the phase space of the Hamiltonian
formalism, in the subsection 2.1, we introduce a family of
conjugate momenta (\ref{cmntcm}) from the family of Lagrangian
$L_{\ep}$, i.e. $$ p_{j \ep }=\frac {\pa L_{\ep}} {\pa {\dot
{q}^j}_{\ep}}, $$ and taken a Legendre transformation to get a set
of the Hamiltonian functions in the family in (\ref{ltfmcm}), i.e.
$$ H_{\ep} :=H({q^i}_{\ep} ,p_{j \ep })=p_{k \ep } {{\dot
q}^k}_{\ep} -L({q^i}_{\ep}, {{\dot q}^j}_{\ep} ). $$ We have also
calculated the variation of action functional along the direction
$k$ in (\ref{deltaS2}) where the differentiation of a family of
action functionals have been taken as follows \be\label{dSepcm}
dS_{\epsilon}=
\int_{t_1}^{t_2}dt \{(\dot q_{\epsilon} ^i-\frac {\partial
H_{\epsilon} } {\partial {p_{i
\epsilon} }})d p_{i \epsilon} -(\frac {%
\partial H_{\epsilon} } {\partial {\ q_{\epsilon} ^i}}+\dot p_{i
\epsilon}) d q_{\epsilon} ^i+\frac{d}{dt}(p_{i \epsilon} d
q_{\epsilon} ^i)\}. \ee

Let us introduce a pair sets of canonical Euler-Lagrange 1-form
families in the Hamiltonian formalism
as follows: 
\be \label{elfhm1} E_{p \ep } (q_{\ep} ^i, p_{j \ep} )=-( \dot
{p}_{j \ep}  + \frac {\pa H_{\ep} } {\pa {q^j}_{\ep}})d
{q^j}_{\ep},\qquad
 E_{q \ep } ({q^i}_{\ep}, p_{j
\ep} )=({{\dot q}^j}_{\ep}-
 \frac {\pa H_{\ep} } {\pa p_{j \ep} } )dp_{j \ep}.
\ee And a family of the canonical 1-forms \be
\theta_{H \epsilon}=p_{i \epsilon} d q_{\epsilon} ^i. \ee Thus,
the equation (\ref{dSepcm}) can be expressed as
\be\label{dSepcm2} dS_{\epsilon}=
\int_{t_1}^{t_2}dt \{E_{p \ep }+E_{q \ep } +\frac{d}{dt}\theta_{H
\epsilon}\}. \ee This equation can be viewed as the equation for
1-forms.

Furthermore, due to the nilpotency of $d$ in the multi-parameter
differential approach, it is straightforward from \be\label{ddShm}
0=d^2 S_{\epsilon}|_{\epsilon^k}=0 \ee to
get\be\label{dEhm1}d(E_{p
 }+E_{q })+\frac{d}{dt}\omega_{H}=0. \ee Here $\omega_H$ is the
symplectic 2-form in the Hamiltonian formalism \be
d\omega_H=d\theta_H =dp_i \wedge dq^i.\ee

 We may introduce a family of $ z_{\ep}^T=(p_{\ep} ^T,
q_{\ep} ^T)$, where $p_{\ep} ^T =(p_{1 \ep } ,\cdots, p_{n \ep} ),
q_{\ep} ^T=({q^1}_{\ep},\cdots, {q^n}_{\ep}),$  defined by \be
z_{\ep}(t) :=z(t)+{\ep^l} \d_l z(t), \ee where \be \d_l z(t)=\f d
{d\ep^l}\mid_{\ep^l=0}z_{\ep}(t)\ee
infinitesimal variation of $z(t)$ along the direction $l$ in the
configuration space.

Then the Euler-Lagrange 1-forms in (\ref{elfhm1}) become \be
\label{elfhm2} E_{z \ep} (z_{\ep} , \dot z_{\ep} )=dz_{\ep}
^T(\nabla_{z_{\ep} } H_{\ep}-J \dot z_{\ep}), \ee where $J$ is a
$2n \times 2n$ symplectic matrix.

Taking the exterior differential of $E_{z \ep}$ and setting
$\epsilon^k=0$ afterwards, it follows that
\be\label{dEhm2}dE_{z}+\frac{1}{2}dz^T \wedge J dz=0.\ee This is
equivalent to the equation (\ref{dEhm1}).

\subsubsection{Remarks on Euler-Lagrange cohomology and symplectic structure
preserving property in classical mechanics}

$\quad$ It is easy to check that under the Legendre transformation
(\ref{ltfmcm}) all relevant issues in the Lagrangian and
Hamiltonian formalism are in one-to-one correspondence.
 For instance, we have the following equivalent
relations:\be\begin{array}{rcl} Lagrangian~ formalism \quad &&
\quad Hamiltonian ~formalism\\
L(q^i, {\dot q}^i) \quad &\Leftrightarrow& \quad
p_i{\dot q}^i-H(p_i, q^j)\\
E( q^i, {\dot q}^i) \quad
&\Leftrightarrow& \quad E_p(p_i,
q^j)+E_q(p_i, q^j)\\
\theta_L \quad &\Leftrightarrow& \quad \theta_H\\
dL=E+\frac{d}{dt}\theta_L \quad &\Leftrightarrow& \quad d(p_i{\dot
q}^i-H)=E_p+E_q+\frac{d}{dt}\theta_H\\
\omega_L \quad &\Leftrightarrow& \quad \omega_H\\
 dE+\frac{d}{dt}\omega_L=0 \quad
&\Leftrightarrow& \quad d(E_p+E_q)+\frac{d}{dt}\omega_H=0
\end{array} \ee

Therefore, 
these important issues on the Euler-Lagrange cohomology and the
symplectic structure preserving property
in two formalisms are corresponding to each other. 
Let us enumerate these issues in the Hamiltonian formalism.

First, the null canonical Euler-Lagrange 1-forms with ${\ep} ^l=0$
give rise to a pair  sets of the canonical equations of motion
(\ref{ceqs1}). 
In terms of $z$, the null canonical Euler-Lagrange 1-form
(\ref{elfhm2}) with ${\ep} ^l=0$ gives rise to the canonical
equations in $z(t)$: \be \label{hen2} \dot z=J^{-1}\nabla_z H. \ee

Secondly, the above null forms are the special case of the
coboundary canonical Euler-Lagrange 1-forms, say, \be E(z, \dot
z)=d\a(z, \dot z), \ee where $\a(z, \dot z)$ is an arbitrary
function of $(z, \dot z)$ on the phase space.

Thirdly,  from either the expression of $d(p_i{\dot q}^i-H)$ or 
the definitions of the canonical Euler-Lagrange 1-forms it is easy
to see that the canonical Euler-Lagrange 1-forms are not exact in
general. Therefore, there exists a nontrivial Euler-Lagrange
cohomology in the Hamiltonian formalism for classical mechanics
defined as 
\vskip 2mm {\centerline {${ H_{CM}}$:=\{Closed Euler-Lagrange
forms\}/ \{Exact Euler-Lagrange forms\}.} } \vskip 2mm

 Fourthly, from the equation (\ref{dEhm1}) or (\ref{dEhm2}) 
it follows a theorem on the symplectic structure preserving law in
the Hamiltonian mechanics.

{\it{\it Theorem 3.1}:

On the phase space of the Hamiltonian mechanics, the symplectic
structure $\om_H $ preserving law \be \f d {dt} \om_H  =0 \ee
holds {\it if and only if} the canonical Euler-Lagrange forms are
closed: \be d E(z, \dot z)=0, \quad i.e. \quad d(E_p(q^i,
p_j)+E_q(q^i, p_j))=0. \ee}

This means  that the symplectic conservation law holds not only in
the solution space of the equations of motion as shown in the
standard approach \c{ALD78}\c{AM78} but also in the function space
with the closed Euler-Lagrange condition \c{GLW01}\c{hyg1}.

Finally, it should be mentioned that 
the Euler-Lagrange cohomological scenario in the Hamiltonian
formalism may be performed in two slightly different processes.
Namely, it may either start from the exterior derivative of the
action functional or begin with the canonical equations. In the
second process, the families of the canonical Euler-Lagrange
1-forms (\ref {elfhm1}) and (\ref{elfhm2}) may be introduced
directly referring to the corresponding canonical equations
(\ref{ceqs1}) and (\ref{hen2}) but releasing first all canonical
variables from the solution space of the canonical equations
(\ref{ceqs1}) and (\ref{hen2}) respectively. Then by taking the
exterior differential of the families of the canonical
Euler-Lagrange 1-forms (\ref{elfhm1}) and (\ref{elfhm2})
and setting the free parameters being vanish afterwards it also
follows the theorem 3.1 as was shown in (\ref{dEhm2}).

\subsection{Discrete Euler-Lagrange cohomology and symplectic structure
preserving property in difference discrete classical 
mechanics}

$\quad$

\subsubsection{
Difference discrete 
Lagrangian formalism}

$\quad$ Let us first  consider the discrete Euler-Lagrange
cohomology, its relation to the simplectic structure and its
preserving property for the difference discrete classical
 mechanics in the Lagrangian formalism.

The difference discrete Lagrangian at the moment $t_k$ is written
as in (\ref{LD1}) on $F(T{\cal M}_k)$. That is
$$
{L_D}^{(k)}=L_D(q^{i (k)}, q_t^{i (k)}),
$$
where ${q_t}^{i (k)}$ is the forward difference of $q^{i (k)}$
defined in (\ref{fd1}).

 Taking the exterior differential $d$ of $
{L_D}^{(k)}$ in the function space $F(T{\cal M}_k)$  as has been
done in the subsection 2.1 in the framework of multi-parameter
differential approach, 
it follows
$$ d{L_D}^{(k)}= \frac {\pa {L_D}^{(k)}} {\pa q^{i (k)}}
dq^{i (k)}
           +\frac {\pa {L_D}^{(k)}} {\pa {q_t}^{i (k)}} d {q_t}^{i (k)}.
$$
Using the modified Leibniz law (\ref{lbnz2}) with respect to the
forward difference $\Delta_t=\pa_t$ defined in (\ref{fd1}) and
introducing the discrete Euler-Lagrange  1-form as well as the
discrete canonical 1-form
 $\theta_{D L}^{(k)}$
\be \label{delf1} {E_D}^{(k)}(q^{i (k)}, {q_t}^{j (k)}):=\{\frac
{\pa {L_D}^{(k)}} {\pa q^{i (k)}}
           -\Delta_t(\frac {\pa {L_D}^{(k-1)}} {\pa (\Delta_t q^{i(k-1)})})\}dq^{i(k)},
\ee 
\be {\theta_{D L}}^{(k)}=\frac {\pa {L_D}^{(k-1)}} {\pa (\Delta_t
q^{i (k-1)})}dq^{i (k)}, \ee
 we have
\be \label{dL03} d{L_D}^{(k)}={E_D}^{(k)}+
           \Delta_t{\theta_D}^{(k)}.
\ee

Due to the nilpotency of $d$ on  $T^*({\cal M}_k)$,
$d^2{L_D}^{(k)}=0$, we get \be d {E_D}^{(k)}+\Delta_t {\om_{D
L}}^{(k)}=0, \ee where ${\om_{D L}}^{(k)}$ is a discrete
symplectic 2-form on $T^*({\cal M}_k)$ \be {\om_{D
L}}^{(k)}=d{\theta_{D L}^{(k)}} =\frac {\pa^2 {L_D}^{(k-1)}} {\pa
q^{i (k)} \pa(\Delta_t q^{j (k-1)})} dq^{i (k)}
 \wedge dq^{j (k)}
+\frac {\pa^2 {L_D}^{(k-1)}} {{\pa (\Delta_t q^{i (k)})} {\pa
(\Delta_t q^{j (k-1)})}} d {q_t} ^{i (k)}
 \wedge d q^{j (k)}.\ee

\subsubsection{Difference discrete Hamiltonian
mechanics}

$\quad$ Let us now study the symplectic structure preserving
property in Hamiltonian formalism for the difference discrete
mechanics when the time $t$ is discretized in the manner of last
subsection. 
We start with what has been
constructed for the
 difference discrete mechanics in Lagrangian formalism. 

Let us first define the discrete canonical conjugate momentum \be
{p_i}^{(k)}=\frac {\pa {L_D}^{(k-1)}} {\pa ({\Delta_t q^{i
(k-1)}})}. \ee Now the difference discrete Hamiltonian can be
introduced through the discrete Legendre transformation
\be\label{dlt1} {H_D}^{(k)}={p_i}^{(k+1)}\Delta_t q^{i (k)}-{L_D}
^{(k)}. \ee

A set of canonical equations for the time difference discrete
derivative of $p_i^{(k)}$ follow from the difference discrete
Euler-Lagrange equations (\ref{del1})  and the above discrete
Legendre transformation (\ref{dlt1}) \be \Delta_t
{p_i}^{(k)}=-\frac {\pa {H_D}^{(k)}} {\pa q^{i (k)}}. \ee

From the discrete Legendre transformation, another set of
canonical equations for the time difference discrete derivative of
$q^{i (k)}$ follow \be \Delta_t q^{i (k)}=\frac {\pa {H_D}^{(k)}}
{\pa
{p_i}^{(k+1)}}, \ee 

In terms of $z^{(k)}$, a pair of the canonical equations
(\ref{dHE1}) become \be\label{dHE2} \Delta_t
z^{(k)}=J^{-1}\nabla_z {H_D}^{(k)}(z^{(k)}). \ee

We now consider the relevant cohomological issues in the
difference discrete version of the Hamiltonian mechanics. This
time we start from  introducing the discrete Euler-Lagrange
1-forms. To this end, all time discrete canonical variables $(q^{i
(k)}, {p_j}^{(k)})$ should be released from the solution space of
the difference discrete canonical equations (\ref{dHE1}) or
(\ref{dHE2}). This can easily be realized by means of the
multi-parameter differential approach. Thus a pair of discrete
Euler-Lagrange 1-forms can be introduced: \be\label{dHEE1} {E_{D p
}}^{(k)}(q^{i (k)}, {p_j}^{(k)}) =( \Delta_t {p_j}^{(k)} + \frac
{\pa {H_D}^{(k)}} {\pa q^{j (k)}})d q^{j (k)},\\
{E_{D q}}^{(k)}(q^{i (k)}, {p_j}^{(k)}) =( \Delta_t q^{j (k)}
-\frac {\pa {H_D}^{(k)}} {\pa {p_j}^{(k)}})d {p_j}^{(k)}, \ee or
in term of $z^{(k)}$ \be\label{dHEE2} {E_{D z
}}^{(k)}(z^{(k)})=d{z^{(k)T}}(J \Delta_t z^{(k)} -\nabla_z
{H_D}^{(k)}(z^{(k)})). \ee

By taking the exterior differential of the discrete Euler-Lagrange
1-forms (\ref{dHEE1}) and (\ref{dHEE2}), it is straightforward to
prove the following formula \be \label{dEhmd}d{E_{D z
}}^{(k)}(z^{(k)})+\Delta_t {\omega_{D H}}^{(k)}=0,\ee where
${\omega_{D H}}^{(k)}$ is the difference discrete  version of the
symplectic 2-form at the moment $t_k$ given by \be {\omega_{D
H}}^{(k)}= \frac{1}{2}d{z^{(k)}}^T \wedge J d z^{(k)}.\ee

We can also start from the exterior differential of
${L_D}^{(k)}={p_i}^{(k+1)}\Delta_t q^{i (k)}-{H_D} ^{(k)}$, 
 introduce the discrete Euler-Lagrange 1-forms and the canonical
1-form then take the second exterior differential to get  above
equations.
\subsubsection{Remarks on discrete Euler-Lagrange cohomology and symplectic
structure preserving property in classical mechanics}

$\quad$ As in the case of continuous classical mechanics, it is
easy to check that under the difference discrete  version of the
Legendre transformation (\ref{dlt1}) all relevant issues in the
difference discrete Lagrangian and difference discrete Hamiltonian
formalism are in one-to-one correspondence. In fact, we have the
following equivalent relations:\be\begin{array}{rcl} Discrete ~
Lagrangian~ formalism  &&
 Discrete ~ Hamiltonian ~formalism\\
{L_D}(q^{i (k)}, {\dot q}^{i (k)})  &\Leftrightarrow&
{p_i}^{(k+1)}{\dot q}^{i (k)}-{H_D}({p_i}^{(k)}, q^{j (k)})\\
E_D( q^{i (k)}, {\dot q}^{i (k)})  &\Leftrightarrow&  E_{D
p}({p_i}^{(k)},
q^{j (k)})+E_{D q} ({p_i}^{(k)}, q^{j (k)})\\
{\theta_{D L}}^{(k)}  &\Leftrightarrow& {\theta_{D H}}^{(k)}\\
d{L_D}^{(k)}={E_D}^{(k)}+\frac{d}{dt}{\theta_{D L}}^{(k)} \quad
&\Leftrightarrow& d(p_i^{(k+1)} {\dot
q}^{i (k)}-H_D^{(k)})=E_{D p}^{(k)}+E_{D q}^{(k)}+\frac{d}{dt}\theta_{D H}^{(k)}\\
{\omega_{D L}}^{(k)}  &\Leftrightarrow&  {\omega_{D H}}^{(k)}\\
 d{E_D}^{(k)}+\frac{d}{dt}{\omega_{D L}}^{(k)}=0
&\Leftrightarrow&  d({E_{D p}}^{(k)}+{E_{D
q}}^{(k)})+\frac{d}{dt}{\omega_{D H}}^{(k)}=0
\end{array} \ee

Therefore, for enumerating the important issues on the difference
discrete  Euler-Lagrange cohomology and the difference discrete
symplectic structure preserving property, we may also work with
one formalism,
the corresponding issues in another
formalism can easily be established. 
Let us also enumerate these issues in the difference discrete
Hamiltonian formalism as follows.

First, the null discrete canonical Euler-Lagrange forms give rise
to the canonical equations and they are the special case of the
coboundary discrete canonical Euler-Lagrange forms.

Secondly, since the first terms in the definitions (\ref{dHEE1})
and (\ref{dHEE2}) are not exact in general so that the discrete
canonical Euler-Lagrange forms are not always exact. Therefore,
there exists a nontrivial difference discrete version of the
Euler-Lagrange cohomology in discrete Hamiltonian mechanics:
\vskip 2mm {\centerline {${ H_{DCM}}$:=\{Closed discrete
Euler-Lagrange
forms\}/\{Exact discrete Euler-Lagrange forms\}.} } \vskip 2mm%

Thirdly, from the equation (\ref{dEhmd}) 
it follows straightforwardly the following theorem for the
necessary and sufficient condition of the difference discrete
symplectic structure preserving law.

{\it Theorem 3.2:

The difference discrete symplectic structure preserving equation
\be \Delta_t {\om_{D H}}^{(k)} =0, \quad i.e. \quad {\om_{D H}
}^{(k+1)}={\om_{D H}}^{(k)} \ee 
holds {\it if and only if} the discrete Euler-Lagrange forms are
closed: \be d{E_{D z}}^{(k)}(z^{(k)})=0. \ee}

Fourthly, the difference discrete symplectic structure preserving
law holds in the function space associated with the difference
discrete version of the closed Euler-Lagrange condition in general
rather than in the solution space of the canonical equations only.

Finally, all these issues can be reached by taking the exterior
differential of ${p_i}^{(k+1)}{\dot q}^{i (k)}-{H_D}^{(k)}$ first,
then introducing the discrete Euler-Lagrange 1-forms and discrete
canonical 1-form. The theorem 3.2 follows from
$d^2({p_i}^{(k+1)}{\dot q}^{i (k)}-{H_D}^{(k)})=0.$ In fact, in
the difference discrete  version of the Lagrangian formalism, it
has  started and progressed in this manner.


\section{Euler-Lagrange cohomology, multisymplectic structure preserving
property in continuous and difference discrete classical field
theory}

$\quad $ We now study the discrete Euler-Lagrange cohomology and
its relation to the discrete multisymplectic structure in discrete
classical field theory. In order to self-contained, we first
recall the the content on Euler-Lagrange cohomology and
multisymplectic structure preserving property in continuous
 classical field theory \c{GLW01}\c{hyg1} as well as the multi-parameter
 differential approach
 to the exterior differential in the function space
 in the subsection 4.1 for the both Lagrangian and Hamiltonian formalism.
 For the sake of simplicity, for the difference discrete cases, let us
consider the 1+1-d and 2-d cases in discrete classical field
theory for the a set of generic fields $u^{\a}$ without
constraints in the Lagrangian and Hamiltonian formalism
respectively in the subsection 4.2.

\subsection{ Euler-Lagrange cohomology, multisymplectic structure preserving
property in continuous  classical field theory}

$\quad$ We recall some content on the Euler-Lagrange cohomology
and its relation to the  multisymplectic structure preserving
property in both Lagrangian and Hamiltonian formalism for
continuous classical field theory by means of the multi-parameter
differential approach.

\subsubsection{
Classical field theory in Lagrangian formalism}

$\quad$ In the multi-parameter differential approach to the
variation of the action of the classical field theory in
subsection 2.2.1, the variation along the direction $\beta$ of the
action has been given by the equation (\ref{voa2}), i.e.
$$\d_\beta S:=\f \partial {\partial \ep^{\b}}\mid_{\ep^{\b}=0}
S_{\ep}.
$$
It ia shown that the differentiation of the action functional with
respect to the free parameters $\ep^{\b}$ is given by
$$
d S_{\ep} =\int d^nx\{(\f {\pa \cal L_{\ep}} {\pa u_{\ep}^{\a}}
-\pa_{\mu}({ \f {\pa \cal L_{\ep}} {\pa u_{\mu \ep}^{\a}}})) d
u_{\ep}^{\a} +\pa_{\mu}(\f {\pa \cal L} {\pa u_{\mu \ep}^{\a}} d
u_{\ep}^{\a})\}. $$
Note that the integrand in the above equation
is an equation for 1-forms: \be \label{dL04} d{\cal L}_{\ep}=(\f
{\pa \cal L_{\ep}} {\pa u_{\ep}^{\a}}
-\pa_{\mu}({ \f {\pa \cal L_{\ep}} {\pa u_{\mu \ep}^{\a}}}))
du_{\ep}^i +\pa_{\mu}(\f {\pa \cal L} {\pa u_{\mu \ep}^{\a}}
du_{\ep}^{\a}). \ee

Let us define a family of the Euler-Lagrange 1-forms \be
\label{elf2} E_{\ep}( u_{\ep}^{\a}, u_{\mu \ep}^{\a}):= (\f {\pa
{\cal L}_{\ep}} {\pa u_{\ep}^{\a}}
-\pa_{\mu}({ \f {\pa \cal L_{\ep}} {\pa u_{\mu \ep}^{\a}}}))
du_{\ep}^{\a}, \ee and $n$ sets of 1-forms that each set
corresponds to a family of canonical 1-forms \be
\theta_{\ep}^{\mu}:={ \f {\pa \cal L_{\ep}} {\pa u_{\mu
\ep}^{\a}}}du_{\ep}^{\a}. \ee Then the equation (\ref{dL04})
becomes \be \label{dL05} d{\cal L}_{\ep}=E_{\ep}( u_{\ep}^{\a},
u_{\mu \ep}^{\a})+\pa_{\mu}\theta_{\ep}^{\mu}. \ee

 Furthermore, due to the nilpotency of $d$ with respect
to $\ep^k$,
 taking the second exterior differential of ${\cal L}_{\ep}(u_{\ep}^{\a},
 {u_{\mu \ep}^{\a}})$
 and setting $\ep^{\b}=0$
afterwards $ d^2{\cal L}_{\ep}(u_{\ep}^{\a}, {u_{\mu
\ep}^{\a}})\mid_{\ep^{\b}=0}=0,
$
it follows that \be\label{dEowftL} dE(u^{\a}, u_{\mu}^{\a}) +\frac
{\pa} {\pa {x^{\mu}}} {\om_L}^{\mu} =0, \ee where  ${\om_L}^{\mu}$
are $n$ symplectic structures defined by \be {\omega_L}^{\mu} =
d\theta^{\mu} = \frac {\pa^2 \cal L} {\pa { u^{\a}} {\pa
u_{\mu}^{\b}}} du^{\a} \we du^{\b} +\frac {\pa^2 \cal L} {\pa {
u_{\nu}^{\a}}{\pa {u_{\mu}^{\b}}}} du_{\nu}^{\a} \we du^{\b}. \ee
And they do not change if the set of $n$ canonical 1-forms
transform as \be {\theta^{\mu}} \rightarrow
{\theta^{\mu}}'={\theta^{\mu}}+d\b(u^{\a}, u_{\mu}^{\a}), \ee
where $\b(u^{\a}, u_{\mu}^{\a})$ is an arbitrary function of
$(u^{\a}, u_{\mu}^{\a})$.

\subsubsection{
Classical field theory in Hamiltonian formalism}

$\quad $ All steps that have been progressed in 
the Lagrangian field theory 
can also be well progressed very similarly  in the
Hamiltonian formalism. In order to do so, 
we first  define a set of ``momenta" that are canonically
conjugate to the field variables $$
\pi_{\b}(x)=\frac {\pa {\cal L}} {\pa \dot {u}^{\b}}, $$
and take a Legendre transformation (\ref{ltfmft}) to get the
Hamiltonian density $$
{\cal H}(u^{\a},\pi_{\a},
\nabla_a u^{\a})=\pi_{\a}(x) {\dot u}^{\a}(x) -
{\cal L}(u^{\a}, {\dot u}^{\a}, \nabla_a u^{\a}), ~ a=1,\cdots,n-1. $$

Let us consider the action $S(u^{\a})$
becoming a family of functionals as follows \be S\rightarrow
S_{\ep}=
\int d^{n}x\{\pi_{\a \epsilon}(x) {\dot u_{\ep}}^{\a}(x) - {\cal
H}_{\ep}(u_{\ep}^{\a}, \pi_{\a \epsilon}, 
\nabla_a u_{\ep}^{\a})\}.\ee The differential of $S_{\ep}$ can be
calculated in the multi-parameter differential approach and the
integrand's differential $d{\cal L}_{\epsilon}$  reads
\be\label{dLftH}\begin{array}{l} d{\cal
L}_{\epsilon}|_{\epsilon^{\beta}=0}=d{\pi_{\a}}({\dot {u}^{\a
}}-\frac{\partial{{\cal H}}}{\partial {\pi_{\a}}})-({\dot
\pi_{\a}} +\frac {\pa {\cal H }} {\pa u^{\a}}-\nabla_a(\frac {\pa
{\cal H }} {\pa (\nabla_a u^{\a (i,j-1)})})) d u^{\a (i,j)}\\[4mm]
    \qquad\qquad+\nabla_t({\pi_{\a}} d u^{\a
})-\nabla_a(\frac {\pa {\cal H}} {\pa (\nabla_a u^{\a })}
        d u^{\a}).\end{array}\ee

Introducing a pair sets of the canonical Euler-Lagrange 1-forms
in the Hamiltonian formalism \be \label{elfth}
E_{u}=d\pi_{\a}\{{\dot u}^{\a}(x)-\f {\pa \cal H} {\pa {\pi_{\a}
(x)}}\},\quad E_{\pi}=du^{\a} \{ -\dot {\pi}_{\a}(x)-\f {\pa \cal
H} {\pa {u^{\a} (x)}} +\nabla_a {\f {\pa \cal H} {\pa ({\nabla_a
{u^{\a} (x)}})}} \},\ee and $n$-canonical 1-forms
\be\label{cfmftH} \theta^0=\pi_{\a} d u^{\a }, \quad \theta^{a
}=\frac {\pa {\cal H}} {\pa (\nabla_a u^{\a })}
        d u^{\a },\ee
then we have \be\label{elfmthetafth}d{\cal
L}_{\epsilon}|_{\epsilon^{\beta}=0}=E_{u}+E_{\pi}
+\nabla_t{\theta}^0-\nabla_a{\theta}^{a }.\ee

Due to the nilpotency of $d$, $d^2{\cal
L}_{\epsilon}|_{\epsilon^{\beta}=0}=0$, it is straightforward to
get the following formula: \be\label{dEowftH}
d(E_{u}+E_\pi)+\nabla_t{\om^0}-\nabla_a{\om^a}=0, \ee where
${\om^0}$ and ${\om^a}$ a set of $n$ symplectic 2-forms \be
\om^0=d{\pi_{\a}} \we du^{\a},\qquad \om^a= d(\f {\pa {\cal H}}
{\pa ({\nabla_a {u^{\a} (x)}})}) \we du^{\a}. \ee

\subsubsection{Remarks on Euler-Lagrange cohomology and multisymplectic
structure preserving for classical field theory}

$\quad$ It is easy to find that, similar to the case of classical
mechanics, under the Legendre transformation (\ref{ltfmft}) all
relevant issues in the Lagrangian and Hamiltonian formalism are in
one-to-one correspondent equivalent. In fact, we have also the
following equivalent relations:\be\begin{array}{rcl} Lagrangian~
formalism \quad &&
\quad Hamiltonian ~formalism\\
{\cal L}(u^{\alpha}, { u_{\mu}}^{\alpha}) \quad &\Leftrightarrow&
\quad
\pi_{\alpha}{\dot u}^{\alpha}-{\cal H}\\
E( u^{\alpha}, { u_{\mu}}^{\alpha}) \quad &\Leftrightarrow& \quad
E_{u}+E_{\pi}\\
{\theta_L}^{\mu} \quad &\Leftrightarrow& \quad {\theta_H}^0, {\theta_H}^a \\
d{\cal L}=E+\partial_{\mu}{\theta_L}^{\mu} \quad &\Leftrightarrow&
\quad d(\pi_{\alpha}{\dot u}^{\alpha}
-{\cal H})=E_{u}+E_{\pi}+\partial_{\mu}{\theta_H}^{\mu}\\
{\omega_L}^{\mu} \quad &\Leftrightarrow& \quad {\omega_H}^0, {\omega_H}^a\\
 dE+\partial_{\mu}{\omega_L}^{\mu}=0 \quad
&\Leftrightarrow& \quad
d(E_u+E_\pi)+\partial_{\mu}{\omega_H}^{\mu}=0
\end{array} \ee

Therefore, for enumerating the important issues on the
Euler-Lagrange cohomology and the multisymplectic structure
preserving property in classical field theory, we may work with
one formalism,
then the corresponding issues in another 
formalism are indicated automatically.
Let us this time enumerate and verify these relevant 
issues in the Lagrangian formalism.

First, if the Euler-Lagrange 1-form in (\ref{elf2}) is null with
$\ep^{\b}=0$, i.e. \be E_{\ep}( u_{\ep}^{\a}, u_{\mu
\ep}^{\a})\mid_{\ep^{\b}=0}=0, \ee it gives rise to the
Euler-Lagrange equations (\ref{elenft}).

Secondly, $E_{\ep}( u_{\ep}^{\a}, u_{\mu \ep}^{\a})=0$ is a
special case of the coboundary Euler-Lagrange 1-forms \be E_{\ep}(
u_{\ep}^{\a}, u_{\mu \ep}^{\a})=d\a_{\ep}( u_{\ep}^{\a}, u_{\mu
\ep}^{\a}), \ee where $\a_{\ep}( u_{\ep}^{\a}, u_{\mu \ep}^{\a})$
a family of arbitrary functions of $( u_{\ep}^{\a}, u_{\mu
\ep}^{\a})$. Although they are cohomologically trivial but it can
already be seen that in the Euler-Lagrange 1-forms, $(u^{\a},
u_{\mu}^{\a})$ are already  not in the solution space of the
Euler-Lagrange equations only rather they are in the function
space with corresponding closed Euler-Lagrange condition (see
below) in general.

Thirdly, if the Lagrangian density $\cal L$ in (\ref{Lft}) changes
to $\cal L'$ by adding certain term \be {\cal L}(u^{\a},
u_{\mu}^{\a}) \rightarrow {\cal L}'(u^{\a}, u_{\mu}^{\a})  = {\cal
L}(u^{\a}, u_{\mu}^{\a}) +{\cal V} (u^{\a}), \ee where ${\cal V}
(u^{\a}) $ is an arbitrary function of $u^{\a}$, the equation
(\ref{dL04}) changes to \be\label{dL05} d{\cal
L}'_{\ep}\mid_{\ep^{\b}=0}=E'(u^{\a}, u_{\mu}^{\a}) +\frac {\pa}
{\pa x^{\mu}} {\theta^{\mu}}, \ee where $E'(u^{\a}, u_{\mu}^{\a})$
differs from $E(u^{\a}, u_{\mu}^{\a})$  by changing ${\cal L}$ to
${\cal L}'$ in the expressions, while a set of $n$ canonical
1-forms $\theta^{\mu}$ have not been changed because ${\cal
V}(u^{\a})$ does not depend on $u_{\mu}^{\a}$. In fact, the
Euler-Lagrange equations have been changed by adding a
potential-like term that does not depend on $u_{\mu}^{\a}$. This
means that even if by adding a coboundary
 term,  the Euler-Lagrange equations do change and the set of canonical forms
may  still be the same as before. Furthermore, the canonical
transformations or the multisymplectic mappings that preserve the
Euler-Lagrange equations, correspondingly the canonical equations
in the Hamiltonian formalism, and Euler-Lagrange 1-forms as well
as the multi-symplectic structures {\it form} invariant. This will
lead to the issues on the generating functions, its relation to
the Euler-Lagrange cohomology and so on. We will explore these
issues elsewhere.

Fourthly,  from the equation (\ref{dL04}) it is easy to see that
$E_{\ep}( u_{\ep}^{\a}, u_{\mu \ep}^{\a})$ in general are not
cohomologically trivial because the families of canonical 1-forms
are not trivial. Therefore, There exists a nontrivial
Euler-Lagrange cohomology in the classical field theory for the
set of generic fields $u^{\a}(x)$:

\vskip 2mm {\centerline {$H_{CFT}$:=\{ closed Euler-Lagrange
forms\}/\{ exact Euler-Lagrange forms\}.}}

\vskip 2mm
 Furthermore, 
From the equations (\ref{dEowftL}) and (\ref{dEowftH}) in the
Hamiltonian formalism it follows  straightforwardly an important
theorem in the classical Lagrangian  field theory.

{\it Theorem 4.1:

For a given Lagrangian field theory, there exists a set of $n$
symplectic structures ${\om_L}^{\mu}$ and the  multisymplectic
preserving property,
 i.e. the conservation or divergence free  law of the multisymplectic structures
\be \frac {\pa} {\pa {x^{\mu}} }{\om_L}^{\mu} =0 \ee holds {\it if
and only if } the relevant Euler-Lagrange 1-form is closed \be d
E(u^{\a}, u_{\mu}^{\a})=0. \ee}

It is interesting to see that if we introduce a new 2-form \be
{\Om}(u^{\a}, u_{\mu}^{\a})=d E(u^{\a}, u_{\mu}^{\a}). \ee Then
$\Om$ may be viewed as a $U(1)$-like curvature 2-form while the
Euler-Lagrange 1-form the $U(1)$-like connection 1-form.
Therefore, the closed Euler-Lagrange condition is nothing but the
flat connection condition. On the other hand, if for some reason
that the multisymplectic conservation law is broken then the
broken pattern may be described by the curvature 2-form $\Om$.
There is a similar issue in the finite dimensional case as well.

It is also important to notice that the multisymplectic structure
preserving  property is directly linked with the closed
Euler-Lagrange condition. And although the null Euler-Lagrange
1-form, the coboundary Euler-Lagrange 1-forms satisfy the
Euler-Lagrange condition, it does not mean that the closed
Euler-Lagrange 1-forms can always be exact as was pointed out
above. In addition, $u^{\a}(x)$'s in the Euler-Lagrange condition
are {\it not} in the solution space of the Euler-Lagrange
equations only in general. Therefore, the  multisymplectic
structure preserving  property, i.e. the conservation law of the
set of $n$
 symplectic 2-forms
$\omega^{\mu}$, holds not only in the solution space of the field
equations but also in the function space with the closed
Euler-Lagrange condition in general.

Analog to the case of classical mechanics, there is also an
another slightly different way to deal with the issues on the
Euler-Lagrange cohomology and the multisymplectic structure
preserving property. Namely, either taking the exterior
differential of the Lagrangian first, or directly starting from
the Euler-Lagrange equations (\ref{elenft}). By means of the
multi-parameter differential approach, it is easy to release all
field variables form the solution space of the Euler-Lagrange
equations to the function space $F(TM)$ and introduce a family of
the Euler-Lagrange 1-forms (\ref{elf2}) associated with the
equations (\ref{elenft}). Then by taking the exterior differential
of the Euler-Lagrange 1-forms and setting $\ep^{\a}=0$, it is
straightforwardly to re-derive the theorem on the necessary and
sufficient condition for the multisymplectic structure preserving
law.

\subsection{Discrete Euler-Lagrange cohomology and
multisymplectic structure preserving property in discrete field
theory}

$\quad$ For the sake of simplicity, we consider the cases of 1+1
or $2$  dimensional base manifold.  Let $X^{1,1}$ or $X^2$ with
suitable signature of the metrics be the base manifold, $L^2$ a
regular lattice with 2-directions $x_{\mu},(\mu=1, 2)$  on
$X^{1,1}$ or $X^2$, $M_D$ the configuration space with $u^{\a
(i,j)} \in M_D$ and so forth as before.

\subsubsection{
Discrete classical field theory in Lagrangian formalism}

$\quad$ The difference discrete Lagrangian for a set of the
generic fields $u^{\a}, \a=1, \cdots, r,$ is a functional in
$F(T({{\cal M}_{{\cal X}_{(i,j)}}}))$ \be {{\cal
L}_D}^{(i,j)}={\cal L}_D(u^{\a (i,j)}, {\Delta_{\mu} u}^{\a
(i,j)}). \ee Taking exterior differential $d \in T^*({{\cal
M}_{{\cal X}_{(i,j)}}})$ of ${{\cal L}_D}^{(i,j)}$ and making use
of the modified Leibniz law (\ref{lbnz2}), in the framework of
multi-parameter differential approach, we get \be \label{dL06}
d{{\cal L}_D}^{(i,j)}= 
           E_D
           (u^{\a (i,j)}, {\Delta_{\mu} u}^{\a (i,j)})+\Delta_{\mu} {\theta_{D L}}^{\mu (i,j)},
\ee where ${E_D}^{(i,j)}$ are the discrete Euler-Lagrange 1-forms
defined by \be {E_D}
(u^{\a (i,j)}, {\Delta_{\mu} u}^{\a (i,j)}):=\{\frac {\pa {\cal
L}_D^{(i,j)}} {\pa u^{\a (k,l)}} -\Delta_1 (\frac {\pa {\cal
L}_D^{(i-1,j)}} {\pa ({\Delta_1 u}^{\a (k-1,l)})}) -\Delta_2(\frac
{\pa {\cal L}_D^{(i,j-1)}} {\pa ({\Delta_2 u}^{\a (k,l-1)})}) \}d
u^{\a (k,l)}, \ee
and ${\theta_{D L}}^{\mu (i,j)}$ are two canonical 1-forms: \be
{\theta_{D L}}^{1 (i,j)}=\frac {\pa {\cal L}_D^{(i-1,j)}} {\pa
({\Delta_1 u }^{\a (k-1,l)})} du^{\a (k,l)},\quad \quad {\theta_{D
L}}^{2 (i,j)}=\frac {\pa {\cal L}_D^{(i,j-1)}} {\pa ({\Delta_2
u}^{\a (k,l-1)})} du^{\a (k,l)}. \ee It is easy to see that there
exist two symplectic 2-forms on $T^*({{\cal M}_{{\cal
X}_{(i,j)}}})$: \be {\om_{D L}}^{\mu (i,j)}=d{\theta_{D L}}^{\mu
(i,j)},\quad \mu=1,2. \ee The equation $d^2{{\cal
L}_D}^{(i,j)}=0,$
 on  $T^*({{\cal M}_{{\cal X}_{(i,j)}}})$ leads to the discrete multisymplectic
structure preserving  property, i.e. the conservation law or the divergence
free equation
 of $\om^{\mu (i,j)}$: \be\label{dEowftD} dE_D(u^{\a (i,j)}, {\Delta_{\mu} u}^{\a (i,j)})
+\Delta_{\mu} {\omega_{D L}}^{\mu (i,j)}=0.\ee

\subsubsection{
Discrete classical field theory in Hamiltonian formalism}

$\quad$ Let us now study  the difference discrete classical field
theory in Hamiltonian formalism in the case of the spacetime/space
$x^{\mu}, \mu=1,2,$ are discretized in the manner of last
subsection while the configuration space at each node  of the
relevant lattice $L^2$ with coordinates $x(i,j), (i,j) \in Z
\times Z$ and the ones on its neighboring are still continuous.

In order to transfer to the difference discrete  version of the
Hamiltonian formalism, we first define a set of the discrete
canonical conjugate momenta on the tangent space of the set of
nodes neighboring to the node $x(i,j)$, i.e. on $T({\cal
X}_{(i,j)})$:
 \be {\pi_{\a}}^{
(i,j)}=\frac {\pa {{\cal L}_D}^{(i-1,j)}} {\pa ({\Delta_t {u}^{\a
(i-1,j)})}}. \ee The time difference discrete derivative of
$\pi^{\a (i,j)}$ follow from the discrete Euler-Lagrange equation
(\ref{del2}) \be \Delta_t {\pi_{\a}}^{ (i,j)}=\frac {\pa {{\cal
L}_D}^{(i,j)}} {\pa {u}^{\a (i,j)}} -\Delta_x ( \frac {\pa {{\cal
L}_D}^{(i,j-1)}} {\pa (\Delta_x{u}^{\a (i,j-1)})}). \ee Now the
difference discrete Hamiltonian is introduced through
the discrete Legendre transformation \be {{\cal H}_D}^{(i,j)}
={\pi_{\a}}^{(i+1,j)}{\Delta_t {u}^{\a (i,j)}}-{{\cal
L}_D}^{(i,j)}. \ee

The difference discrete  version of action functional is given by
\be {S_{D H}}=\sum_{(i,j)}{{\cal L}_{D
}}^{(i,j)}=\sum_{(i,j)}\{{\pi_{\a}}^{(i+1,j)}{\Delta_t {u}^{\a
(i,j)}}-{{\cal H}_D}^{(i,j)}.\} \ee The differential of each term
can be calculated in the multi-parameter differential approach
\be\label{dLftHd}\begin{array}{l} d{{\cal L}_{D
\epsilon}}^{(i,j)}|_{\epsilon^{\beta}=0}=d{\pi_{\a}}^{(i+1,j)}({\Delta_t
{u}^{\a (i,j)}}-\frac{\partial{{\cal H}_D}^{(i,j)}}{\partial
{\pi_{\a}}^{(i+1,j)}})-({\Delta_t \pi_{\a}}^{(i,j)} +\frac {\pa
{\cal
H }_D^{(i,j)}} {\pa u^{\a (i,j)}}\\[2mm]-\Delta_x(\frac {\pa {\cal H
}_D^{(i,j-1)}} {\pa (\Delta_x u^{\a (i,j-1)})})) d u^{\a (i,j)}
    +\Delta_t({\pi_{\a}}^{(i,j)} d u^{\a
(i,j)})-\Delta_x(\frac {\pa {\cal H}_D^{(i,j-1)}} {\pa (\Delta_x
u^{\a (i,j-1)})}
        d u^{\a (i,j)})\end{array}\ee

Now we introduce the difference discrete  version of the canonical
Euler-Lagrange 1-forms \be\label{elfmftHD}\begin{array}{l}{E_{D
u}}^{(i,j)}=d{\pi_{\a}}^{(i+1,j)}({\Delta_t {u}^{\a
(i,j)}}-\frac{\partial{{\cal H}_D}^{(i,j)}}{\partial
{\pi_{\a}}^{(i+1,j)}}),\\[2mm]
{E_{D \pi}}^{(i,j)}=d u^{\a (i,j)}(-{\Delta_t \pi_{\a}}^{(i,j)}
-\frac {\pa {\cal
H }_D^{(i,j)}} {\pa u^{\a (i,j)}}
+\Delta_x(\frac {\pa {\cal H
}_D^{(i,j-1)}} {\pa (\Delta_x u^{\a (i,j-1)})})) 
,\end{array}\ee and the difference discrete  version of two
canonical 1-forms \be\label{cfmftHD} {\theta_{D H}}^{0
(i,j)}={\pi_{\a}}^{(i,j)} d u^{\a (i,j)},\quad {\theta_{D H}}^{1
(i,j)}=\frac {\pa {\cal H}_D^{(i,j-1)}} {\pa (\Delta_x u^{\a
(i,j-1)})}
        d u^{\a (i,j)}.
\ee Thus the equation (\ref{dLftHd}) becomes \be\label{EthetaftHD}
d{{\cal L}_{D \epsilon}}^{(i,j)}|_{\epsilon^{\beta}=0}={E_{D
u}}^{(i,j)}+{E_{D \pi}}^{(i,j)}+\Delta_t{\theta_{D H}}^{0
(i,j)}-\Delta_x{\theta_{D H}}^{1 (i,j)}.\ee

Now due to the nilpotency of $d$, $d^2{{\cal L}_{D
\epsilon}}^{(i,j)}|_{\epsilon^{\beta}=0}=0$, it follows the
equation for difference discrete multisymplectic structure
preserving property: \be\label{dEowftHD} d{E_{D u}}^{(i,j)}+d{E_{D
\pi}}^{(i,j)}+\Delta_t{\omega_{D H}}^{0 (i,j)}-\Delta_x{\omega_{D
H}}^{1 (i,j)}=0, \ee where ${\omega_{D H}}^{0 (i,j)}$ and
${\omega_{D H}}^{1 (i,j)}$ are two symplectic 2-forms: \be
{\omega_{D H}}^{0 (i,j)}= d{\pi_{\a}}^{(i,j)}\wedge d u^{\a
(i,j)},\quad{\omega_{D H}}^{1 (i,j)}=d\frac {\pa {\cal
H}_D^{(i,j-1)}} {\pa (\Delta_x u^{\a (i,j-1)})}\wedge
        d u^{\a (i,j)}.\ee

\subsubsection{Remarks on discrete Euler-Lagrange cohomology and
multisymplectic structure preserving property in discrete field
theory}

$\quad$ It is clear that similar to the case for the classical
field theory we have also the following equivalent relations
between the difference discrete  versions of the Lagrangian and
Hamiltonian formalism for the difference discrete  classical field
theory:\be\begin{array}{rcl}
Discrete ~Lagrangian~ formalism  && Discrete~ Hamiltonian ~formalism\\
{{\cal L}_D}^{(i,j)}
 &\Leftrightarrow&
{\pi_{\alpha}}^{(i+1,j)}{\Delta_t {u}^{\a (i,j)}}-{{\cal H}_D}^{(i,j)}\\
{E_D}^{(i,j)}
&\Leftrightarrow&
{E_{D \pi}}^{(i,j)}+{E_{D u}}^{(i,j)}\\
{\theta_{L D}}^{\mu {(i,j)}} &\Leftrightarrow& {\theta_{H D}}^{0 (i,j)}, {\theta_{H D}}^{a (i,j)} \\
d{{\cal L}_D}^{(i,j)}={E_L}^{(i,j)}+\Delta_{\mu}{\theta_{L
D}}^{\mu {(i,j)}}
 &\Leftrightarrow&  d(\pi_{\alpha}{\dot u}^{\alpha}
-{\cal H})^{(i,j)}=(E_{D u}+E_{D \pi}+\Delta_{\mu}{\theta_{D H}}^{\mu})^{(i,j)}\\
{\omega_{L D}}^{\mu {(i,j)}} &\Leftrightarrow&
{\omega_{H D}}^{0 (i,j)}, {\omega_{H D}}^{a (i,j)}\\
 d{E_D}^{(i,j)}+\Delta_{\mu}{\omega_{L D}}^{\mu {(i,j)}}=0
&\Leftrightarrow& d(E_u+E_\pi)_D^{(i,j)}+\Delta_{\mu}{\omega_{H
D}}^{\mu {(i,j)}}=0
\end{array} \ee


Let us  enumerate and verify relevant important issues on the
discrete Euler-Lagrange cohomology and difference discrete
multisymplectic structure preserving property in the difference
discrete Lagrangian formalism.


First, the null discrete Euler-Lagrange 1-form  corresponds to the
discrete Euler-Lagrange equations and it is a special case of
coboundary discrete Euler-Lagrange 1-forms \be
 {E_D}^{(i,j)}=d{\a_D}^{(i,j)},
\ee where ${\a_D}^{(i,j)}$ an arbitrary function on $F(T^*{\cal
M}_{{\cal X}_{(i,j)}})$.

Secondly, although they satisfy the discrete Euler-Lagrange
condition, it does not mean that all closed discrete
Euler-Lagrange 1-forms are exact. As  a matter of fact, from the
equation (\ref{dL06}) it is easy to see that the Euler-Lagrange
1-forms are not exact in general since the two canonical 1-forms
${\theta_D}^{\mu (i,j)}, (\mu=1,2)$ are not
trivial. Therefore, 
for a given difference discrete  field theory, there exists a
nontrivial difference discrete version of the Euler-Lagrange
cohomology:

\vskip 2mm {\centerline {$H_{DCFT}$:=\{closed Euler-Lagrange
forms\}/\{exact Euler-Lagrange forms\}.}}

\vskip 2mm

Thirdly, from the equations (\ref{dEowftD}) and (\ref{dEowftHD})
it follows the theorem on the necessary and sufficient condition
for the difference discrete  multisymplectic structure preserving
law:

{\it Theorem 4.2:

The difference discrete multisymplectic structure preserving  law
\be \Delta_{\mu}\om_D^{\mu(i,j)}=0 \ee holds {\it if and only if}
the discrete Euler-Lagrange 1-form satisfies the discrete
Euler-Lagrange condition, i.e. it is closed: \be d
{E_D}^{(i,j)}=0.\ee}

 In addition, this also indicates that the
variables $u^{\a (k,l)}$'s etc. in the cohomology are still in the
function space rather than the ones in the solution space only.
Consequently, this means that the difference discrete
multisymplectic structure preserving law holds in the function
space with the closed discrete Euler-Lagrange condition in general
rather than in the solution space only.


Finally, it should be mentioned that 
all these issues
can be straightforward to generalize to higher dimensional cases
of spacetime $X^{1,n-1}$ or  space $X^{n}$ as base manifold.



\section{Difference discrete variational principle and discrete Euler-Lagrange cohomological approach to
symplectic and multisymplectic algorithms}

$\quad$ It is worthwhile to show that the difference discrete
variational principle and the cohomological scenario described in
the previous sections for the difference discrete mechanics  as
well as for the discrete classical field theory in the both
Lagrangian and Hamiltonian formalism can be directly applied to
the numerical schemes in the symplectic and multisymplectic
algorithms respectively.

As a matter of fact, the difference discrete versions of the
Euler-Lagrange equations and the canonical equations of motion in the both Lagrangian 
and Hamiltonian formalism offer themselves the numerical schemes
in symplectic and multisymplectic algorithms for the difference
discrete version of the classical mechanics and 
the difference discrete version of the classical field theory
respectively.

In addition, as has been pointed out in \c{GLWW01} that the
Euler-Lagrange-like cohomological approach may also be applied to
the so-called Hamiltonian-like ODEs and PDEs respectively.
Therefore, for the difference discrete versions of these
Hamiltonian-like ODEs and PDEs the difference discrete variational
principle and the difference discrete version of
the 
cohomological approach may also be
available.

 We investigate the relevant issues for the symplectic
algorithm first in the subsection 5.1, and the relevant issues in
the multisymplectic algorithm first in the subsection 5.2.

\subsection{Difference discrete variational principle and discrete  cohomological approach to symplectic  algorithm}

$\quad$ As in the case of both  Lagrangian and Hamiltonian
formalism for classical mechanics, the above-established
difference discrete variational principle and the discrete
cohomological scenario in the difference discrete Lagrangian and
Hamiltonian formalism for classical mechanics should be directly
applied to the difference discrete versions of ODE's with
Lagrangian and/or Hamiltonian and to the numerical schemes for the
symplectic algorithm. In this section we study this topic. We
first consider how to apply the difference discrete variational
principle to the numerical schemes in symplectic algorithm in the
subsection 5.1.1. Then we show the difference discrete version of
the Euler-Lagrange cohomological scenario offers how to justify a
scheme is
symplectic in subsection 5.1.2. 

\subsubsection{Difference discrete variational principle and symplectic  algorithm}

$\quad$ Let us first note that both  
the difference discrete versions of the Euler-Lagrange equations
(\ref{del1}) in the Lagrangian formalism and the difference
discrete  versions of the canonical equations (\ref{dHE1}) or
(\ref{dHE2}) in the Hamiltonian mechanics are derived from a
relevant difference discrete variational principle in the
framework of the multiparameter differential approach. These
difference discrete equations may in fact offer certain numerical
schemes and  are automatically symplectic.

In the difference discrete  Lagrangian formalism case, the
discrete Euler-Lagrange equations (\ref{del1}) read
$$
\frac {\pa {L_D}^{(k)}} {\pa q^{i (k)}}
           -\Delta_t(\frac {\pa {L_D}^{(k-1)}} {\pa ({\Delta_t q^{i
           (k-1)}})})=0.
$$
Their offering a numerical scheme that is symplectic can be seen
manifestly from the difference discrete Euler-Lagrange equations
(\ref{del2}) in the example 2.1:
$$
\Delta_t(\Delta_t q^{i (k-1)})-\f {\pa} {\pa q} V(q^{i (k)})=0,
$$ i.e. $$ \frac 1 {\t ^2} (q^{i (k+1)}-2q^{i (k)}+q^{i (k-1)})
=\f {\pa} {\pa q} V(q^{i (k)}).
$$
For given initial values of $q^{i (k=0)}, q^{i (k=1)}$, the
equations give rise to $q^{i (k=2)}$ and so on. For the Lagrangian
(\ref{lexmple1}) this is in fact a simplest scheme  that
preserves, in the sense of time difference discrete, the
symplectic structure. This has been proved in the subsection 3.1.

For the case of difference discrete Hamiltonian mechanics , the
difference discrete canonical equations in the (\ref{dHE1}) or
(\ref{dHE2}) read
$$\Delta_t q^{i (k)}={{p_i}^{(k+1)}},\qquad \Delta_t {p_i}^{(k)}=-\f {\pa}
{\pa q} V(q^{i (k)}),$$ and $$ \Delta_t z^{(k)}=J^{-1}\nabla_z
{H_D}^{(k)}(z^{(k)}).$$ They also offer a set of numerical schemes
in difference discrete Hamiltonian formalism and preserve the
symplectic structure in the sense of time difference discrete.
This has been proved in the subsection 3.1 as well.

In fact, in the multi-parameter differential approach to
difference discrete variational principle, it is easy to see why
the difference discrete variational principle offers the numerical
schemes that are automatically symplectic. It has been shown that
the variations of the difference discrete action functionals are
calculated by taking the differential of the difference discrete
action functionals and setting $\epsilon^k=0$ afterwards, i.e.
$dL_{D \epsilon}|_{\epsilon^k=0}$ in the Lagrangian formalism and
$
d(p_i\Delta_t q^i-H)_{D \epsilon}|_{\epsilon^k=0}$ in the
Hamiltonian formalism. As was shown in the subsections 2.1 and
3.1, the stationary requirement of difference discrete variational
principle leads to either difference discrete Euler-Lagrange
equations or difference discrete canonical equations of motion.
Furthermore, the second exterior differentials of the difference
discrete action functionals lead to the symplectic structure
preserving law.

Let us consider the Euler mid-poind scheme. We find that there
exists a difference discrete action functional for the scheme. 

The difference discrete Lagrangian of the scheme can be given by
\be\label{Lmdpt} {L_{mdpt}}^{(k)}=L(q^{i (k+\frac{1}{2})},
\Delta_t q^{i (k)}), \quad q^{i (k+\frac{1}{2})}:=\frac{1}{2}(q^{i
(k+1)}+q^{i (k)}).\ee Note that at the moment $t_k$ $q^{i
(k+\frac{1}{2})}$ and $\Delta_t q^{i (k)}$ are the coordinates and
the tangents for the scheme.

Then the discrete canonical momenta conjugated to the coordinates
can be defined as \be\label{pinmdpt}
{p_i}^{(k+\frac{1}{2})}=\frac{\partial
{L_{mdpt}}^{(k)}}{\partial(\Delta_t q^{i (k)})}, \quad
{p_i}^{(k+\frac{1}{2})}:=\frac{1}{2}(p_i^{(k+1)}+p_i^{(k)}).\ee
The discrete Legendre transformation is given by \be H(q^{i
(k+\frac{1}{2})},
{p_i}^{(k+\frac{1}{2})})={p_i}^{(k+\frac{1}{2})}\Delta_tq^{i
(k)}-{L_{mdpt}}^{(k)}.\ee
The difference discrete action functional of the scheme is given
by \be\label{action functionalmdpt} S_{mdpt}=\sum_{k \in
Z}\{p_i^{(k+1/2)}\Delta_t q^{i (k)}-H(q^{i (k+1/2)},
p_i^{(k+1/2)}\}. \ee

Let us take the  variation of this difference discrete action
functional in the framework of the multi-parameter differential
approach, i.e. the variation along the direction $l$ is
manipulated by \be
\delta_lS_{mdpt}=\frac{\partial}{\partial\epsilon^l}S_{mdpt
\epsilon}|_{\epsilon^l=0}.\ee Eventually, the differential of each
term under the summation of $S_{mdpt}$  is given by
\be\label{dLmdpt}\begin{array}{l}d{L_{mdpt}}^{(k)}:=d
{p_i}^{(k+1/2)}(\Delta_tq^{i (k)}-\frac{\partial}{\partial
p}H(q^{i (k+1/2)}, {p_i}^{(k+1/2)}))\\[4mm]\quad -(\Delta_t
{p_i}^{(k)}+\frac{\partial}{\partial p}H(q^{i (k+1/2)},
{p_i}^{(k+1/2)}))d q^{i (k+{1/2})}+\Delta_t({p_i}^{(k)}dq^{i
(k)}).\end{array}\ee Note that here  the following generalized
modified Leibniz law \c{GLWW02} (see also the appendix) has been
used:\be\label{lbnzgm}\begin{array}{l} \Delta_t(f^{(k)} \cdot
g^{(k)}):=\frac{1}{\tau}(f^{(k+1)} \cdot
g^{(k+1)}-f^{(k)} \cdot g^{(k)})\\[4mm]\qquad\qquad =\frac{1}{\tau}(\Delta_t f^{(k)} \cdot
g^{(k+1/2)}+f^{(k+1/2)} \cdot \Delta_t g^{(k)}). \end{array}\ee
The first equality in the above equation is the definition for the
forward difference, while the second can be easily proved.

 It is now
clear that the stationary requirement of the difference discrete
variational principle leads to the Euler mid-point scheme as
follows:\be\label{mdpt1}\begin{array}{l}\Delta_tq^{i
(k)}=\frac{\partial}{\partial p}H(q^{i (k+1/2)},
{p_i}^{(k+1/2)}),\\[4mm]\Delta_t
{p_i}^{(k)}=-\frac{\partial}{\partial p}H(q^{i (k+1/2)},
{p_i}^{(k+1/2)}).\end{array}\ee

Thus we have shown that the Euler mid-point scheme is a difference
discrete variational  scheme with corresponding  (dependent)
difference discrete variables and suitable Leibniz law for the
difference.

\subsubsection{Discrete 
cohomological approach to symplectic  algorithm}





$\quad$ We now consider how to apply the difference discrete
version of the Euler-Lagrange cohomological approach to the
numerical schemes in symplectic algorithm.

 In the standard approach, it
 is commonly accustomed to
regarding a numerical scheme as a (time-discrete) mapping. In
order to justify whether a given numerical scheme is symplectic,
the standard approach 
is to see whether this mapping is symplectic preserving and the
verification is always carried out in the solution space of the
scheme \c{KF84}\c{SC94}. In the difference discrete version of the
Euler-Lagrange cohomological approach, however, instead of working
on the solution space, it is working on the function space with
the relevant cohomological issues. Analog to the case for the
difference discrete classical mechanics, there are two slightly
different ways to apply the cohomological approach. Namely, one is
based upon the difference discrete variational principle for the
schemes and taking second (exterior) differential of the action
functional to get the necessary and sufficient condition for the
symplectic structure preserving property of the scheme. Another is
 to release the scheme away from the solution space, even if the
solution space does exist, and to introduce some suitable
difference discrete Euler-Lagrange 1-forms associated with the
scheme such that the null difference discrete Euler-Lagrange
1-forms give rise to the scheme. Then by taking the exterior
derivative of the difference discrete Euler-Lagrange 1-forms to
see whether follows a time-discrete symplectic structure
preserving law.

We will consider some examples to show how the cohomological
scenario works.  Let us first retain to the mid-point scheme
(\ref{mdpt1}) and follow-up by the cohomological approach.

It is clear that by introducing the discrete Euler-Lagrange
1-forms \be\label{elfmmdpt}\begin{array}{l}{E_q}^{(k)}=d
{p_i}^{(k+1/2)}(\Delta_tq^{i (k)}-\frac{\partial}{\partial
p}H(q^{i (k+1/2)}, {p_i}^{(k+1/2)})),\\[4mm]{E_p}^{(k)}= -(\Delta_t
{p_i}^{(k)}+\frac{\partial}{\partial p}H(q^{i (k+1/2)},
{p_i}^{(k+1/2)}))d q^{i (k+{1/2})},
\end{array}\ee
and the difference discrete  canonical 1-form for the scheme:\be
{\theta_{mdpt}}^{(k)}={p_i}^{(k)}dq^{i (k)},\ee the equation
(\ref{dLmdpt}) becomes \be\label{dLmdpt2}
d{L_{mdpt}}^{(k)}={E_q}^{(k)}+{E_p}^{(k)}+\Delta_t
{\theta_{mdpt}}^{(k)}.\ee
 By taking exterior
differential of the above equation (\ref{dLmdpt2}), due to
$d^2L_{mdpt}=0$,
it follows that \be\label{thmmdpt} d(E_q+E_p)^{(k)}+\Delta_t
{\omega_{mdpt}}^{(k)}=0,\ee where
${\omega_{mdpt}}^{(k)}=d{p_i}^{(k)}\wedge dq^{i (k)}$. Since  the
null Euler-Lagrange 1-forms of (\ref{elfmmdpt}) give rise to the
scheme and automatically satisfy $d(E_q+E_p)^{(k)}=0$, while the
latter leads to the symplectic conservation law from
(\ref{thmmdpt}):
$$\Delta_t
{\omega_{mdpt}}^{(k)}:=\frac{1}{\tau}({\omega_{mdpt}}^{(k+1)}-
{\omega_{mdpt}}^{(k)})=0.$$ Therefore, the midpoint scheme is
symplectic.

On the other hand, we can also start directly from the  the
mid-point scheme in terms of $z^{(k)}$:
 \be \Delta_t z^{(k)}=J^{-1}\nabla_z
H(\f 1 2 (z^{(k+1)}+z^{(k)})). \ee This time we release the scheme
form the solution space first, even if it does exist. This can be
treated by the multi-parameter differential approach. For
simplicity, we suppose this has been done already. Then we
introduce a difference discrete
Euler-Lagrange 1-form for the scheme 
\be {E_{z mdpt}}^{(k)}=\f 1 2 d({z^{(k+1)}+z^{(k)})}^T\{ J
\Delta_t z^{(k)} -\nabla_z {H}(\f 1 2 (z^{(k+1)}+z^{(k)}))\} \ee
such that the null discrete Euler-Lagrange form gives rise to the
scheme. Then by taking the exterior differential of ${E_{z
mdpt}}^{(k)}$
 in the function space, it follows
$$
d{E_{z mdpt}}^{(k)}=\f 1 2 d({z^{(k+1)}+z^{(k)})}^T\wedge J
\Delta_t
dz^{(k)}. $$
Therefore, the difference discrete symplectic structure preserving
law \be \Delta_t (d{z^{(k)}}^T\wedge J dz^{(k)})=0 \ee holds {\it
if and only if} the discrete Euler-Lagrange form is closed: \be d
{E_{z mdpt}}^{(k)}=0. \ee

Furthermore, due to the exact forms do not change the closed
condition, the null form may be redefined by adding certain exact
forms so that the scheme may be generalized to a type of schemes
while the difference discrete symplectic structure preserving  law
is  the same.

It is also interesting that this issue offers a way to generalize
the scheme to the high order ones while the difference discrete
symplectic structure preserving law is the same.

Let us consider the 4-th order symplectic scheme as follows
\c{FW89}:
\be\begin{array}{l}
\nabla_t z^{(n)} = J^{-1}\nabla_z H(\frac{1}{2}(z^{(n+1)}+z^{(n)}))\quad\quad\\[4mm]
\quad\quad-\frac{h^2}{24}J^{-1}\nabla_z((\nabla_zH)^TJH_{zz}J\nabla_zH)
(\frac{1}{2}(z^{(n+1)}+ z^{(n)})).
\end{array}\ee
Introduce a new ``Hamiltonian" ${\cal H} $ 
\begin{eqnarray}
{\cal H} = H-\frac{h^2}{24}(\nabla_zH)^TJH_{zz}J\nabla_zH,
\end{eqnarray}
 then this 4th-order symplectic scheme becomes
\begin{eqnarray}
\nabla_t z^{(n)} = J^{-1}\nabla_z{\cal H}(\frac{1}{2}(z^{(n+1)}+
z^{(n)})).
\end{eqnarray}
The  discrete Euler-Lagrange 1-form associated with this case now
can be introduced: \be {E_{z 4th}}^{(k)}=\f 1 2
d({z^{(k+1)}+z^{(k)})}^T\{ J \Delta_t z^{(k)} -\nabla_z {\cal
H}(\f 1 2 (z^{(k+1)}+z^{(k)}))\}. \ee It is easy to check that
these two discrete Euler-Lagrange forms differ by an exact form:
\be
{E_{z mdpt}}^{(k)}-{E_{z 4th}}^{(k)}=\frac{h^2}{24}d\a,
\ee where $\a=(\nabla_zH)^TJH_{zz}J\nabla_zH$ is a function of $\f
1 2 (z^{(k+1)}+z^{(k)})$.

This means that
they are cohomologically equivalent. In addition, this also
indicates that the 4-th order midpoint scheme is difference
discrete variational as well.

\subsection{Difference discrete variational principle and discrete cohomological approach to multisymplectic algorithm}

$\quad$ As in the case of both Lagrangian and Hamiltonian
formalism for classical field theory, the scenario of the
above-established difference discrete variational principle and
cohomological approach to the difference discrete Lagrangian and
Hamiltonian formalism for classical field theory should be
directly applied to the difference discrete versions of PDEs with
Lagrangian and/or Hamiltonian and to the numerical schemes for the
multisymplectic algorithm.

\subsubsection{Difference discrete variational principle and multisymplectic algorithm}

$\quad$ In this part of the subsection, we first point out that
the difference discrete  field equations derived by the difference
discrete variational principle themselves offer numerical schemes
for the multisymplectic algorithm. We also derive a new scheme for
the difference discrete Hamiltonian
formalism by the difference discrete variational principle with the generalized modified Leibniz law (\ref{lbnzgm}).
Then we
study how to apply the difference discrete variational principle
approach to the numerical schemes for
 the Hamiltonian-like PDEs named in \c{GLWW01}. A type of so-called
Hamiltonian PDEs proposed first in \c{TB97} (see also \c{SR00})
are in fact a type of the Hamiltonian-like PDEs.

As was just mentioned, the difference discrete Euler-Lagrange
equations (\ref{del3}) derived via the difference discrete
variational principle in the difference discrete Lagrangian
formalism for classical field theory, i.e.
$$ \frac {\pa {L_D}^{(i,j)}} {\pa u^{\a (k,l)}} -\Delta_1 (\frac {\pa
{L_D}^{(i-1,j)}} {\pa u_{1}^{\a (k-1,l)}}) -\Delta_2(\frac {\pa
{L_D}^{(i,j-1)}} {\pa u_{2}^{\a (k,l-1)}}) =0 $$ offer themselves
a numerical scheme that preserves the multisymplectic structures
for the 2-dimensional spacetime/space in difference discrete
version.

For the difference discrete  Hamiltonian formalism of the field
theory, the difference discrete canonical equations (\ref{ceqsfd})
derived via the difference discrete variational principle, i.e.
$$\Delta_t {u}^{\a (i,j)}=\frac {\pa {{\cal H}_D}^{(i,j)}} {\pa
{\pi_{\a}}^{(i+1,j)}}, $$
$$ \Delta_t {\pi_{\a}}^{(i,j)}=-\frac {\pa {{\cal H}_D}^{(i,j)}}
{\pa {u}^{\a (i,j)}} +\Delta_x ( \frac {\pa {{\cal
H}_D}^{(i,j-1)}} {\pa
(\Delta_x{u}^{\a (i,j-1)})})$$ 
also offer themselves a set of numerical schemes for the
multisymplectic algorithm. It has been proved in the subsection
4.2 that they preserve the multisymplectic structures  as well for
 the difference discrete spacetime/space.

In the last subsection, it was shown that by the difference
discrete variational principle with the generalized modified
Leibniz law the Euler midpoint-scheme had been derived from a
difference discrete action functional. Similarly, the same
generalized modified Leibniz law (\ref{lbnzgm}) can be employed to
construct the numerical schemes for the multisymplectic algorithm.
Let us consider the case in the difference discrete  Hamiltonian
formalism 
for the difference discrete field theory.


We  introduce the difference discrete  conjugate momenta%
\begin {eqnarray}
 {\pi_{\alpha}} ^{(i+\frac {1}{2},j)}:=
 \frac{\partial {{\cal L}_D}^{(i,j)}}{\partial (\Delta_t
u^{\alpha (i,j+\frac {1}{2})})},\qquad {\pi_\beta} ^{(i+\frac
{1}{2},j)}= \frac {1}{2} ({\pi_\beta} ^{(i+1,j)} +{\pi_\beta}
^{(i,j)}), 
\end {eqnarray}
where $${{\cal L}_D}^{(i,j)}:={\cal L}_D({u}^{\alpha(i+\frac
{1}{2},j+\frac {1}{2})}, \Delta_{t} u^{\alpha(i,j+\frac {1}{2})},
\Delta_x u^{\alpha(i+\frac {1}{2},j)})$$ is the difference
discrete Lagrangian density with suitable discrete variables
${u}^{\alpha(i+\frac {1}{2},j+\frac {1}{2})}$, $\Delta_{t}
u^{\alpha(i,j+\frac {1}{2})}$, $\Delta_x u^{\alpha(i+\frac
{1}{2},j)}$, and
\begin {eqnarray}\label{ui+1/2} {u} ^{\alpha (i+\frac {1}{2},j)}:=
\frac {1}{2} ({u} ^{\alpha (i+1,j)} +{u} ^{\alpha (i,j)}),\qquad
u^{\alpha (i,j+\frac {1}{2})}:=\frac {1}{2} (u ^{\alpha (i,j+1)}
+u ^{\alpha (i,j)}), \\\nonumber {u}^{\alpha (i+\frac
{1}{2},j+\frac {1}{2})}:=\frac {1}{4}(u^{\alpha
(i+1,j+1)}+u^{\alpha (i,j+1)}+u^{\alpha (i+1,j)}+u^{\alpha
(i,j)}).\quad
\end {eqnarray} The difference
discrete Hamiltonian density via the difference discrete  Legendre
transformation
\begin {eqnarray}\label{boxH}
{{\cal H}_D}^{(i,j)}:= {\cal H}_D({u}^{\alpha(i+\frac
{1}{2},j+\frac {1}{2})}, {\pi_\beta} ^{(i+\frac {1}{2},j)},
\Delta_x u^{\alpha(i+\frac
{1}{2},j)})\qquad\qquad\qquad\qquad\qquad
\\\nonumber ={{\pi_\beta}}^{(i+\frac {1}{2},j)} \Delta_t u^{\alpha(i,j+\frac
{1}{2})} - {\cal L}_D({u}^{\alpha(i+\frac {1}{2},j+\frac {1}{2})},
\Delta_{t} u^{\alpha(i,j+\frac {1}{2})}, \Delta_x
u^{\alpha(i+\frac {1}{2},j)}).
\end {eqnarray}

Now the difference discrete action functional is given by
\begin {eqnarray}
S_D=\Sigma_{i,j \in Z} {{\cal L}_D}^{(i,j)}.
\end {eqnarray}
The variation of $S_D$ can be manipulated in the framework of
multiparameter differential approach and the differential of
${{\cal L}_D}^{(i,j)}$ with respect to $\varepsilon^{\beta}$ can
be calculated to get
   \begin{eqnarray}\label{dLfthdmp}d{{\cal L}_D}^{(i,j)}
    &=&(\Delta_t u^{\beta (i,j+\frac {1}{2})}
    - (\frac {\partial {{\cal H}_D}}{\partial {\pi_\beta}})^{(i,j)})
d{\pi_\beta} ^{(i+\frac {1}{2},j)} \\\nonumber
    & &+(-\Delta_t {\pi_\beta} ^{(i,j)} -
    (\frac {\partial {\cal H}_D}{\partial u^\beta})^{(i,j)}
    +\Delta_x {\pi'_\beta} ^{(i,j)} )
d {u}^{\beta (i+\frac {1}{2},j+\frac {1}{2})}\\\nonumber
    & &+\Delta_t( {\pi_\beta} ^{(i,j)} du^{\beta (i,j+\frac {1}{2})})
    -\Delta_x({\pi'_\beta} ^{(i,j)} d u^{\alpha (i+\frac {1}{2},j)}),
\end{eqnarray}
where  the notation \begin {eqnarray}
   {\pi'_{\beta}}^{(i,j+\frac{1}{2})}=\frac {1}{2} ({\pi'_\beta} ^{(i,j+1)}+ {\pi'_\beta}^{(i,j)})=
   ({\frac {\partial {\cal H}_D}{\partial u^\beta_x}})^{(i,j)}
\end {eqnarray} and the generalized modified Leibniz law (\ref{lbnzgm}) for the
both $\Delta_t$ and $\Delta_x$ have been adopted.

Now the stationary requirement against all variations along any
direction of difference discrete variational principle leads to a
new scheme, which may be called the midpoint box scheme, in the
difference discrete Hamiltonian field theory, with assuming
variations of both $u^{\beta (i,j+\frac {1}{2})}$ and
 $u^{\beta (i+\frac {1}{2},j)}$ vanish at infinity, as follows:
\begin{eqnarray}\label{mptbox}
\Delta_t u^{\beta (i,j+\frac {1}{2})}&=& (\frac {\partial {\cal
H}_D}{\partial {\pi_\beta}})^{(i,j)}\\\nonumber \Delta_t
{\pi_\beta} ^{(i,j)}&=& -  (\frac {\partial {\cal H}_D}{\partial
u^\beta})^{(i,j)}
+\Delta_x {\pi'_\beta} ^{(i,j)},
\end{eqnarray} together with $$
{\pi'_\beta}^{(i,j+\frac {1}{2})}
=(\frac {\partial {\cal H}_D}{\partial u^\beta_x})^{(i,j)}.$$

In fact,  it is shown that this midpoint box scheme can be derived
via the difference discrete variational principle with the
generalized modified Leibniz law (\ref{lbnzgm}).

Finally, let us derive the midpoint box scheme  for a type of PDEs
\c{TB97} by means of the difference discrete variational principle
with the generalized modified Leibniz law (\ref{lbnzgm}):\begin
{eqnarray}\label{tbc} {\bf M}Z_t+\epsilon{\bf K}Z_x=\nabla_z S(Z),
\end {eqnarray} where the same notations in \cite{TB97} have been used and
${\bf K}^T=-{\bf K}, {\bf L}^T=-{\bf L}, \epsilon=\pm$. 

It can be shown \cite{GLWW01} that there is an action functional
for the PDEs (\ref{tbc})
\begin {eqnarray}
S=\int d^2x {\cal L},\qquad\qquad\\\nonumber {\cal L}=\frac {1}{2}
Z^T({\bf M}Z_t +\epsilon{\bf K}Z_x)-S(Z),
\end{eqnarray}where $\cal L$ is the Lagrangian density, and
the PDEs (\ref{tbc}) can be reached by the variational principle
of action functional as the Euler-Lagrange equations.

Let us now consider its difference discrete  version formulation.
First, introduce the difference discrete  Lagrangian density as
follows
\begin {eqnarray}\label{tblgrnd}
{{\cal L}_D}^{(i,j)}={\cal L}_D({Z}^{(i+\frac {1}{2},j+\frac
{1}{2})}, \D_t Z^{(i,j+\frac {1}{2})}, \D_x Z^{(i+\frac
{1}{2},j)}),
\end{eqnarray}%
where 
${Z}^{(i+\frac {1}{2},j+\frac {1}{2})}$ is taken
as the coordinates in the configuration space on the note $(i,j)$
in the difference discrete spacetime, $\D_t Z^{(i,j+\frac
{1}{2})}$ and $\D_x Z^{(i+\frac {1}{2},j)}$ the ones on the
tangent space of the configuration space. In the difference
discrete variational principle, they are regarded as variables to
be variated.
Now the action functional becomes%
\begin {eqnarray}\label{tbactin functionald}
S_D=\sum_{(i,j)\in Z \times Z}{{\cal L}_D}^{(i,j)}
\end {eqnarray}
The variation of $S_D$ along the direction $\beta$ in the
configuration space of the system can be calculated in the
framework of multi-parameter differential approach%
\begin {eqnarray}\label{tbvoAd}
\delta_{\beta}S_D=\frac{\partial}{\partial
\varepsilon^{\beta}}S_{D \varepsilon}|_{\varepsilon^{\beta}=0}.
\end {eqnarray}
It is straightforward to get the differential with respect to
$\varepsilon^{\beta}$ for each term under the summation
\begin {eqnarray} \label{tbdLd}d{{\cal L}_D}^{(i,j)}
   =d{Z}^{(i+\frac {1}{2},j+\frac {1}{2}) T} 
   ( {\bf M} \D_t Z^{(i,j+\frac {1}{2})}
   + {\bf K} \D_x Z^{(i+\frac {1}{2},j)} -\nabla_z
   S({Z}^{(i+\frac {1}{2},j+\frac {1}{2})}))\\\nonumber
   -\frac {1}{2} \D_t (dZ^{(i,j+\frac {1}{2}) T}
   {\bf M} Z^{(i,j+\frac {1}{2})})
   -\frac {1}{2}\epsilon \D_x (dZ^{(i+\frac {1}{2},j) T} 
   {\bf K} Z^{(i+\frac {1}{2},j)}).
   \end {eqnarray}
Here the variables ${Z}^{(i+\frac {1}{2},j+\frac {1}{2})}$, $\D_t
Z^{(i,j+\frac {1}{2})}$ and $\D_x Z^{(i+\frac {1}{2},j)}$ are
regarded as the variational variables, the multi-parameter
$\varepsilon^{\beta}$ are omitted and the generalized modified
Leibniz law (\ref{lbnzgm}) for the differences are adopted.

Then the stationary requirement against all variations along any
direction of difference discrete variational principle leads to
the midpoint box scheme, with assuming variations of both
$Z^{(i,j+\frac {1}{2})}$ and
 $Z^{(i+\frac {1}{2},j)}$ vanish at infinity, as follows: %
\begin {eqnarray}\label{tbbox}{\bf M}\D_t
Z^{(i,j+\frac {1}{2})}+\epsilon{\bf K}\D_x Z^{(i+\frac {1}{2},j)}
=\nabla_z S({Z}^{(i+\frac {1}{2},j+\frac {1}{2})}).
\end {eqnarray}
Thus,  it is proved that the midpoint box scheme for the type of
PDEs in \c{TB97} can also be derived via the difference discrete
variational principle.



\subsubsection{Cohomological approach to
multisymplectic algorithm}

$\quad$ Similar to the case for symplectic algorithm, it should be
emphasized that in the conventional approach to the
multisymplectic algorithm, it
 is commonly accustomed to
regarding a numerical schemes as a set of
(spacetime/space-discrete) mappings. In order to justify whether a
set of given numerical schemes are multisymplectic, the usual
approach to the multisymplectic algorithm is to verify this set of
mappings are multisymplectic preserving and the verification is
always carried out in the solution space of the schemes (see, for
example, \c{TB97}\c{MPS98}). In the difference discrete version of
the Euler-Lagrange cohomological approach, however, instead of
working on the solution space, it is first to release the schemes
away from the solution space, even if the solution space does
exist, and introduce some suitable difference discrete
Euler-Lagrange 1-forms associated with the schemes such that the
null difference discrete Euler-Lagrange 1-forms give rise to the
schemes. Then by taking the exterior derivative of the difference
discrete Euler-Lagrange 1-forms to investigate whether it leads to
a spacetime/space-discrete multisymplectic structure preserving
law.


We first show the multisymplectic property of the midpoint box
scheme (\ref{mptbox}) in the cohomological approach.

 Introducing the difference discrete  Euler-Lagrange 1-forms
\begin{eqnarray}
{E_u}^{(i,j)}:=(\D_t u^{\beta(i,j+\frac {1}{2})}
    - (\frac {\partial {\cal H}_D}{\partial {\pi_\beta}})^{(i,j)})
     d{\pi_\beta} ^{(i+\frac {1}{2},j)},
    \\\nonumber
{E_\pi}^{(i,j)}:=(-\D_t {\pi_\beta} ^{(i,j)} - ( \frac {\partial
{\cal H}_D}{\partial u^\beta})^{(i,j)}
    +\D_x {\pi'_\beta} ^{(i,j)} ) d {u}^{\beta (i+\frac {1}{2},j+\frac {1}{2})},
\end{eqnarray}
and difference discrete  canonical 1-forms
\begin{eqnarray}
\theta^{0 (i,j)}:= {\pi_\beta} ^{(i,j)} du^{\beta (i,j+\frac
{1}{2})},\quad \theta^{1 (i,j)}:={\pi'_\beta} ^{(i,j)} d u^{\beta
(i+\frac {1}{2},j)},
\end{eqnarray}
the equation (\ref{dLfthdmp}) can be rewritten as
\begin{eqnarray}
d{{\cal L}_D}^{(i,j)}={E_u}^{(i,j)}+{E_\pi}^{(i,j)} +\D_t\theta^{0
(i,j)}-\D_x\theta^{1 (i,j)}.
\end {eqnarray}
Taking the exterior differential again, due to the nilpotency of
$d$, $d^2{{\cal L}_D}^{(i,j)}=0$, it follows that
\begin {eqnarray}
d(E_u+E_\pi)^{(i,j)}+\D_t\omega^{0 (i,j)}-\D_x\omega^{1 (i,j)}=0,
\end{eqnarray} where \begin {eqnarray} \omega^{0 (i,j)}:= d{\pi_\beta}
^{(i,j)}\wedge du^ {\beta(i ,j+\frac {1}{2})},\quad\omega^{1
(i,j)}:=d{\pi'_\beta} ^{(i,j)} \wedge d u^{\beta (i+\frac{1}{2},j
)}.
\end{eqnarray}
Thus, the necessary and sufficient condition of the
multisymplectic structure preserving law
\begin {eqnarray}
\D_t\omega^{0 (i,j)}-\D_x\omega^{1 (i,j)}=0,
\end {eqnarray} for the midpoint box scheme is that the associated difference
discrete  Euler-Lagrange 1-forms are closed:
\begin {eqnarray}
d(E_u+E_\pi)^{(i,j)}=0.
\end {eqnarray} Since the null difference discrete  Euler-Lagrange 1-forms
corresponding to the
midpoint box scheme and they do satisfy the closed condition so
that the midpoint box scheme is multisymplectic.

Let us now consider the midpoint box scheme for the type of PDEs
in \c{TB97} and show that it is multisymplectic in the
cohomological approach.

As before, we first define the difference discrete Euler-Lagrange
1-form and two difference discrete canonical 1-forms
\begin {eqnarray}
{E_{box}}^{(i,j)}:=d{Z}^{(i+\frac {1}{2},j+\frac {1}{2}) T} 
( {\bf M} \D_t Z^{(i,j+\frac {1}{2})}
   + \epsilon{\bf K} \D_x Z^{(i+\frac {1}{2},j)} -\nabla_z
   S({Z}^{(i+\frac {1}{2},j+\frac {1}{2})})),
\end {eqnarray}
\begin {eqnarray}
\theta^{0 (i,j)}:=\frac {1}{2} (dZ^{(i,j+\frac {1}{2}) T} 
{\bf M} Z^{(i,j+\frac {1}{2})}), \qquad \theta^{1 (i,j)}:=\frac
{1}{2} (dZ^{(i+\frac {1}{2},j)} \, ^T {\bf K} Z^{(i+\frac
{1}{2},j)})
\end {eqnarray}%
Then the differential of ${{\cal L}_D}^{(i,j)}$ in the equation
(\ref{tbdLd}) can be rewritten as
\begin {eqnarray}
d{{\cal L}_D}^{(i,j)}={E_{box}}^{(i,j)}-\D_t\theta^{0
(i,j)}-\epsilon \D_x\theta^{1 (i,j)}.
\end {eqnarray}
Taking the second exterior differential of the above equation, due
to the nilpotency of $d$, $d^2{{\cal L}_D}^{(i,j)}=0$, it follows
that
\begin {eqnarray}
 0=d{E_{box}}^{(i,j)}-\D_t\omega^{0 (i,j)}-\epsilon
 \D_x\omega^{(i,j)},\end {eqnarray}
where $\omega^{\mu (i,j)}, \mu=0,1$ are two symplectic 2-forms%
\begin {eqnarray}
 \omega^{0 (i,j)}:=-\frac {1}{2} (dZ^{(i,j+\frac {1}{2}) T}
 \wedge {\bf M}
d Z^{(i,j+\frac {1}{2})}),\\\nonumber
\omega^{1 (i,j)}:=-\frac {1}{2}
(dZ^{(i+\frac {1}{2},j) T}
\wedge {\bf K} dZ^{(i+\frac {1}{2},j)}).
\end {eqnarray}
Thus, the necessary and sufficient condition of the
multisymplectic structure preserving law
\begin {eqnarray}
\D_t\omega^{0 (i,j)}-\epsilon
 \D_x\omega^{(i,j)}=0,
\end {eqnarray} for midpoint box scheme of the type of PDEs in
\cite{TB97} is the associated difference discrete  Euler-Lagrange
1-form is closed:
\begin {eqnarray}
d{E_{box}}^{(i,j)}=0.
\end {eqnarray} Since the null difference discrete  Euler-Lagrange 1-form corresponding to the
box scheme and it does satisfy the closed condition so that the
midpoint box scheme for the PDEs in \cite{TB97} is
multisymplectic.

It is clear that for the both midpoint box scheme for Hamiltonian
field theory and  for the type
of Hamiltonian-like PDEs 
the discrete multisymplectic structure preserving law holds in
function space with the discrete closed Euler-Lagrange condition
in general and can also be required in the solution space in each
case.

Finally, it should be mentioned that 
the both above difference discrete variational principle and
cohomological scenario should also be applied to other numerical
schemes in order to justify whether they are variational and
 multisymplectic.

\vskip 2mm
\section {Concluding remarks}

$\quad $
A few remarks are in order:

1. The difference discrete variational formalism widely employed
in this paper is
 different from the one of the Veselov type for the discrete
classical mechanics \c{av88}\c{MV91}. It has been emphasized that
the difference as discrete derivative is an entire geometric
object. The discrete integrants can also combine together  in
certain manner as a geometric object to construct some numerical
schemes as was shown in the section 5. This is more obvious and
natural from the viewpoint of noncommutative geometry. In the
continuous limit, the results given here by the difference
discrete variational principle lead to the correct continuous
counterparts.

It is shown that the difference discrete variational principle
works for the difference discrete  version of classical mechanics
and field theory in the both Lagrangian and Hamiltonian formalisms
that present themselves as symplectic or multisymplectic numerical
schemes and furthermore for other numerical schemes in both
symplectic and multisymplectic algorithms respectively. And the
role-played by the different Leibniz laws for the differences are
quite important in constructing the numerical schemes. 
As a matter of fact,
all numerical schemes in symplectic and multisymlectic algorithms
should be derived by virtue of the difference discrete variational
principle together with the suitable Leibniz law for differences.
We will publish our further investigation on this issue elsewhere
\c{GLWW02}.

2. The cohomological approach adopted in this paper to the
symplectic and multisymplectic geometry and their difference
discrete versions in the both  Lagrangian and Hamiltonian
formalism for the classical mechanics and field theory had been
missed in 
other approaches (see, for example,
\c{ALD78}\c{AM78}\c{KF84}\c{SC94}\c{BSF88}\c{TB97}\c{MPS98}). The
Euler-Lagrange cohomological concepts and their
difference discrete versions, 
such as the Euler-Lagrange 
1-forms,
the null Euler-Lagrange 
1-forms, the
coboundary Euler-Lagrange 
1-forms as
well as the closed Euler-Lagrange 
conditions and their difference discrete  versions, have been
introduced and they have played  very crucial roles in each case
to show that the symplectic and  multisymplectic structure
preserving properties. It has been show that the necessary and
sufficient condition for symplectic and  multisymplectic structure
preserving
property in each case is the related closed Euler-Lagrange condition
being satisfied. 
Therefore, these symplectic and multisymlectic structure
preserving properties hold in the function space with the relevant
Euler-Lagrange condition in general rather than in the solution
space only. Although either the Euler-Lagrange equations and the
canonical equations or the different difference discrete versions
of them do preserve the relevant symplectic and multisymplectic
structures.

It should pointed out that the content of the Euler-Lagrange
cohomology and the roles played by the cohomology in each case
should be further studied not only in classical level but also in
quantum level. And needless to say, this cohomological scenario
should also be tried to apply to other kinds of the mechanics and
field theories such as the ones with different types of
constraints and so on so forth.

3.  As was shown in the text, the variational principle/difference
discrete variational principle and the cohomological approach form
a connecting link between the preceding and the following in
either continuous or difference discrete case. And the
multi-parameter differential approach provides a common framework
for both of them.

 It has been emphasized that  both the
variational principle including  difference discrete variational
principle and the Euler-Lagrange cohomological approach including
its difference discrete version can be directly applied to the
ODEs and PDEs and their discrete versions, which offer themselves
certain numerical schemes in the symplectic and multisymplectic
algorithms, no matter whether there are known Lagrangian and/or
Hamiltonian associated with. In fact, the action functional may be
constructed for certain types of ODEs and  PDEs. Thus,  the
variational principle/difference discrete variational principle
and the scenario of the cohomological approach are also available.

As was emphasized, in the cohomological approach it is always to
release the ODEs, PDEs and numerical schemes away from their
solution spaces and to work on the relevant function space rather
than on the solution space even if it does exist.  In the standard
or conventional approaches to the numerical schemes in symplectic
and multisymplectic algorithms, however, in order to show whether
a given scheme is symplectic or multisymplectic, it is always
working on the solution spaces. The implication of this difference
is quite clear.

4. Some simple noncommutative differential calculus on the regular
lattices are employed in our approach. Since the base space
coordinates $t$ or $x$s are difference discretized and differences
do not satisfy the ordinary commutative Leibniz law for the
differential, in order to study the symplectic and multisymplectic
geometry in these difference discrete systems it is natural and
meaningful to make use of the noncommutative differential
calculus.

5. The  difference discrete version of the classical mechanics and
field theory  can be further generalized. What have been studied
so far is the time discrete classical mechanics and
spacetime/space discrete classical field theory, while the
configuration spaces at each moment of $t_k$ in the discrete
mechanics and the one at each node of the discretized
spacetime/space on the lattice and so on are still continuous. As
a matter of fact, these may be  generalized to the case that the
configuration spaces at each moment and/or node are also
discretized. For the Hamiltonian mechanics, this is closely
related to the case of difference discrete phase space approach to
the systems with separable Hamiltonian \c{GWWWW00}\c{GWZ00}.

6. Finally, it should be mentioned that there are lots of other
problems to be further investigated. \vskip 8mm

\vskip 2mm
\vskip 2mm

{\Large\bf Appendix} \vskip 2mm We have presented the
noncommutative differential calculus on the regular lattice $L^n$
and its applications
 to the discrete symplectic algorithms with  difference discrete
phase space \c{GWWWW00}\c{GWZ00}\c{HYG99}. In this appendix we
briefly recall some content of the noncommutative differential
calculus on $L^n$. General approach to the noncommutative
differential geometry can be found in \c{AC94}.

{\bf {A.1. An noncommutative differential calculus on an Abelian
discrete group}}


Let $G$ be an  Abelian discrete group with a generator $t$, $A$ the  algebra of  complex valued
functions on $G$.

The left and/or  right multiplication of a generator of $G$ on its
element are commute  to each other  since $G$ is  Abelian.  Let us
introduce right action on $A$ that is given by \be R_t f(a)=f(a
\cdot t), \ee where  $f \in  A$,  $a \in  G$, $t$ the generator
and  $\cdot$  the group multiplication.

Let   $V$  be   the   space  of   vector  fields,
  $$V=span\{\partial_t\},$$
where $\partial_t$  is the derivative with respect to the
generator $t$ given by \be (\partial_t f)(a)\equiv R_t f(a)-f(a) =
f(a\cdot t)-f(a). \ee The dual space of $V$, the space of 1-form,
is $\Omega ^1=span \{ \chi^t\}$ that is dual to $V$: \be
\chi^t(\pa_t)=1. \ee The  whole differential  algebra  $\Omega^*$
can also  be defined  as $\Omega^*= \bigoplus\limits_{n=0,1}
\Omega^n $ with $A=\Omega^0$.

Let us define the exterior differentiation in $\Omega^*$ such that
$ d: \Omega^0 \rightarrow \Omega^{1}$. It acts  on a $0$-form $f
\in \omega^0 = A $ is as follows \be df=\partial_tf \chi^l \quad
\in \Omega^{1}. \ee

Now,   the    following   theorem   can    straightforwardly   be   proved.

{\it {\bf Theorem:} The exterior differential operator $d$ is
nilpotent and satisfies \be
\begin{array}{l}
(a)  \quad (df)(v)=v(f), v\in V, f\in \Omega^0,\\
(b) \quad  d(\omega\otimes\omega')=d\omega\otimes\omega'
+(-1)^{deg(\omega)}\omega\otimes d\omega', \quad \om, \om' \in
\Om^*,
\end{array}
\ee
if and only if
\be
\begin{array}{l}
(1) \quad \chi^{i}\otimes  \chi^{j}=(-1)  \chi^{j}  \otimes  \chi^{i},  \\
(2) \quad d\chi^{t}=0 , \\
(3) \quad \chi^{t}f=(R_{t}f) \chi^{t}.
\end{array}
\ee}

This theorem indicates that the $\otimes$-product should be
defined as $\we$-product and $\chi^{t}$, the base of $\Om^1$, can
be denoted as $dt$.

As  was shown  here, in  order to  establish a  well-defined differential
algebra, it  is necessary  and sufficient to  introduce the noncommutative
property   of    the   multiplication   between    function   and   1-form.

The conjugation $ \ast  $ on the whole differential algebra $ \Omega ^{*} $
 and  metric on discrete Abelian group can also be defined.

In the case that the Abelian group is a discrete translation group
with one generator on $R^1$, the action of the group generates a
discrete chain $L^1$ with equal step-length. In the function space
on $L^1$, the integrals can be defined (see, for example,
\c{DMS93}) as follows:
 \be \int_{D -\infty}^{+\infty}df(t):=\sum_{k \in Z}\Delta_t
f(t_k)=f(t_{k=+\infty})-f(t_{k=-\infty}),
\ee

\vskip 2mm

{\bf {A.2. An noncommutative differential calculus on Regular
Lattice} }


Let us consider the discrete translation group $ G^{m}=
\otimes_{i=1}^m G^i$ with $m$ generators, $A$ the function space
on $G^m$ and a regular lattice with equal  spacing in each
direction of $L^m$ on an $m$-dimensional space $R^m$. Here $G^i$
the i-th discrete translation group with one generator acting on
1-dimensional space with coordinate $q$ in such a way: \be
R_{q^i}: q^i_n\rightarrow q^i_{n+1}=q^i_n+h^i, \quad h^i \in R_+,
\ee $R_{q^i}$ the discrete translation operation of the group
$G^i$ and it maps $q^i_n$ of $n$-th node of $q^i$ to the one
$q^i_{n+1}$ at $n+1$-th node, $h^i$ the discrete translation
step-length along this direction and $R_+$ the positive real
number. It is easy to see that  the action of $G^i$ on i-th
1-dimensional  space $R^1$ generates the i-th chain $L^i, ~i=1,
\cdots , m,$ with  equal spacing $h^i$. Similarly, the regular
lattice $L^m$  with equal spacing  $h^i$ on each direction is
generated by $G^m$ acting  on  $R^m$. Since there is a one-to-one
correspondence between nodes on $L^i$ and elements of $G^i$, one
may  simply regard  $L^i$ as  $G^i$. For the same reason, one may
simply regard $L^m$ as $G^m$.

On the nodes of the regular lattice $L^{m}$, there are discrete
coordinates $q^i_n)$, $i=1, \cdots , m$.  There is a set of
  generators in
the  discrete  translation group  $G^{m}$  acting  on $L^{m}$  in  such a  way:
\be
R_{q^i}:  q^{i}_{n} \to q^{i}_{n+1}, \quad i=1, \cdots, m. \ee
With respect to  the generators  there is  a set of independent
 derivatives 
$\pa_{q^i}$ on $f_{n}(q^i)=f(q^i_n) \in  A$. They should be
defined  as the correspondent forward differences of  the
functions valued at two nearest nodes, i.e.
\be
\pa_{q^i}f(q^i_n)=\Delta_{q^i}f(q^i_n)=\f{1}{h}[
(R_{q^i}-id)f(q^i_n)]=\f{1}{h^i}[f(q^i_{n+1})-f(q^i_n)]. \ee
The differential 1-form is defined by
\be
df=
\pa_{q^i} fdq^i=
\Delta_{q^i}fdq^i, \qquad f \in A. \ee The  2-forms  and the whole
differential  algebra $\Omega^*$ can  also be defined. Here $d$ is
the exterior differential operator. Similarly, the following
theorem can be proved for $d$.

{\it {\bf  Theorem}:  $d$ is  nilpotent  and  satisfies the
Leibniz rule,  i.e. \be d^2=0,\quad
d(\om\we\om')=d\om\we\om'+(-1)^{deg(\om)}\om\we d\om', \quad
\om,\om' \in \Omega^*, \ee if and only if \be
f(q^{i}+h^i)dq^{i}=dq^{i}f(q^{i}),\\
q^{i}dq^{i}-dq^{i}q^{i}=-h^idq^{i}.
\ee} The  above two  equations show  the noncommutative properties
between the functions   (including    the   coordinates)    and
differential   forms.

From  these   properties,  it   follows  the  modified   Leibniz
rule  for derivatives (i.e. the forward differences):
\be\label{lbnzm} \Delta_{q^{i}}(f \cdot g)=\Delta_{q^{i}}f \cdot g
+ \{R_{q^{i}}f\} \cdot \Delta_{q^{i}}g.
\ee

It should be noted that first from the definition of the forward
difference, the more general Leibniz law may holds. For example,
 the
generalized modified Leibniz law (\ref{lbnzgm}) used in the
section 5 is the special case of $a=1/2$, while the above modified
Leibniz law (\ref{lbnzm}) is corresponding to $a=1$, in the
following general Leibniz law with an arbitrary parameter $a \in
[0, 1]$ \c{GLWW02}: \be\label{lbnzg}
\begin{array}{l}
\Delta_t(f^{(k)} \cdot
g^{(k)})=\frac{1}{h}\{(af^{(k+1)}+(1-a)f^{(k)})\Delta_tg^{(k)} +
\Delta_tf^{(k)}((1-a)g^{(k+1)}+ag^{(k)})\}|_{a=1/2}\\[4mm]
=\frac{1}{h}(\Delta_t f^{(k)} \cdot g^{(k+1/2)}+f^{(k+1/2)}
\cdot \Delta_t g^{(k)}),\quad f^{(k+1/2)}=1/2(f^{(k+1)}+f^{(k)}). \end{array}\ee%
 It is straightforward to
prove that this general one (\ref{lbnzg}) consists with the
definition for the forward difference. Namely,
$$\Delta_t(f^{(k)}
\cdot g^{(k)}):=\frac{1}{h}(f^{(k+1)} \cdot g^{(k+1)}-f^{(k)}
\cdot g^{(k)}). $$ But the price has to paid is that the
corresponding Leibniz law for the exterior differential operator
$d$ no long holds (except for the case of $a=1$).

Secondly,  the definitions and relations given above for the
noncommutative differential calculus on the regular lattice
$L^{m}$ are at least formally very similar to the ones in the
ordinary commutative differential calculus  on $R^{m}$. The
differences between the two cases are commutative or not.

Similarly, the  contraction between  forms and  differences can
be defined  as the same as the one in $R^{m}$: \be
<dq^i \we dq^j, \Delta_{q^k}>=dq^i \d^j_k,\\
<dq^i \we dq^j, \Delta_{q^k} \Delta_{q^l}>=\d^j_k \d^i_l,\\
i_{X_f}\a=<\a, X_f >, \quad f \in A', \a \in \Omega^1, \ee
where $X_f$ is the  Hamiltonian vector field of $f$. 

The Hodge $*$ operator and the co-differential operator
$$
\delta_L: ~ \Omega^k \rightarrow \Omega^{k-1}
$$
on the regular lattice  $L^{m}$ can also be defined similarly as
the ones on   $R^{m}$ (see, for example, \c{DMS93}). Consequently,
The Laplacian on the lattice $L^{m}$ may also given by \be
\Delta_L=d \delta_L+\delta_L d. \ee It is in fact the discrete
counterpart of the Laplacian $\Delta$ on  $R^{m}$. For other
objects and/or properties on $R^{m}$, there may have the discrete
counterparts on $L^{m}$ as well. For example, the null-divergence
equation of a form $\om $ on $R^m$ reads \be \delta \a =0. \ee Its
counterpart on the lattice $L^m$ is simply \be \delta_L \a_L =0.
\ee This is the right forward difference version of the
divergence-free equation.

The discrete version of the integral on $L^m$ can also be defined
(see, for example, \c{DMS93}).

In the case of $L^{1,m} \in R^{1,m}$ with Lorentz signature, these
equations become the conservation law of $\a$ and its difference
version of $\a_L$. This is available not only for the symplectic
geometry and symplectic algorithms but also the multisymplectic
geometry and multisymplectic algorithms as well. It should be
emphasized that  for the difference discrete counterparts on the
lattice, they obey the noncommutative differential calculus on the
lattice $L^{m}$ rather than the commutative differential calculus
on $R^{m}$. This is the most important point.

\vskip 1.0cm 
{\parindent=0mm\large\bf Acknowledgements} \vskip 0.5cm

 Guo, Li
and Wu would like to thank Professors and/or Drs J. Butcher, Y.B.
Dai,  J.B. Chen, J.L. Hong, A. Iserles,  R.I. McLachlan, D. Lewis,
H. Munthe-Kaas, P. Olver, B.R. Owren, M.Z. Qin, S. Reich, Z.J.
Shang, X.C. Song, G. Sun, Y.F. Tang, L.H. Wang, S.H. Wang, M.
West, Z. Xu for valuable discussions and comments, specially
during the International Workshop on Structure-Preserving
Algorithms, March 25-31, 2001, Beijing. This work is partly
supported by National Science Foundation of China.

\end{document}